\documentclass[review]{elsarticle}
\usepackage{subfigure}
\usepackage{indentfirst}
\usepackage{bm}
\usepackage{threeparttable}
\usepackage{multirow}
\usepackage{amsmath}
\usepackage{array}
\usepackage{amssymb}
\usepackage{graphicx}
\usepackage{url}
\usepackage{hyperref}
\usepackage{booktabs}
\usepackage{color}

\journal{Computers \& Mathematics with Applications}

\begin{document}

\begin{frontmatter}

\title{A simplified finite volume lattice Boltzmann method for simulations of fluid flows from laminar to turbulent regime, Part \uppercase\expandafter{\romannumeral2}: Extension towards turbulent flow simulation}

\author{Yong Wang$^a$}
\ead{wangyong19890513@mail.nwpu.edu.cn}

\author[]{Chengwen Zhong$^a$\corref{mycorrespondingauthor}}
\ead{zhongcw@nwpu.edu.cn}
\author[]{Jun Cao$^b$\corref{mycorrespondingauthor}}
\cortext[mycorrespondingauthor]{Corresponding author}
\ead{jcao@ryerson.ca}

\author[]{Congshan Zhuo$^a$}
\ead{zhuocs@nwpu.edu.cn}
\address{$^a$National Key Laboratory of Science and Technology on Aerodynamic Design and Research, Northwestern Polytechnical University, Xi'an, Shaanxi 710072, China \\
$^b$Department of Mechanical \& Industrial Engineering, Ryerson University, Toronto, Ontario, Canada M5B 2K3}

\begin{abstract}
In this paper, the finite volume lattice Boltzmann method (FVLBM) on unstructured grid presented in Part \uppercase\expandafter{\romannumeral1} of this paper is extended to simulate the turbulent flows. To model the turbulent effect, the $k-\omega$ SST turbulence model is incorporated into the present FVLBM framework and also is solved by the finite volume method. Based on the eddy viscosity hypothesis, the eddy viscosity is computed from the solution of $k-\omega$ SST model, and the total viscosity is modified by adding this eddy viscosity to the laminar (kinematic) viscosity given in the Bhatnagar-Gross-Krook collision term. In order to enhance the computational efficiency, the three-stage second-order implicit-explicit (IMEX) Runge-Kutta method is used for temporal discretization and the time step can be larger one- or two-order of magnitude compared with explicit Euler forward scheme. Though the computational cost is increased, the finial computational efficiency is enhanced about one-order of magnitude and the good results also can be obtained at large time step through the test case of lid-driven cavity flow. Two turbulent flow cases are carried out to validate the present method, including flow over backward-facing step and flow around NACA0012 airfoil. Our numerical results are found to be in agreement with experimental data and numerical solutions, demonstrating applicability of the present FVLBM coupled with $k-\omega$ SST model to accurately predict the incompressible turbulent flows.
\end{abstract}


\end{frontmatter}

\section{Introduction}
As a mesoscopic approach, lattice Boltzmann method (LBM) has received considerable attention since its appearance. The advantages of LBM and its some applications can be found in Ref. \cite{succi2001lattice,guo2013lattice}. In a large number of applications, turbulence is usually encountered as the flow can not maintain laminar state. Due to the complexity of the flow, numerical simulation of turbulent flows is one of most challenge in computational fluid dynamics (CFD). Enormous studies and discusses about turbulence can be found in macro method based on Navier-Stokes equations. As a new CFD method, since its inception, the efforts of using LBM to study and simulate turbulent flows have been unintermittent. For the simulation of turbulent flows, though LBM have some special advantages than macro method\cite{chen2007new}, one of the most important defects is the grid problem. Historically, LBM is originated from the lattice gas automata (LGA)\cite{mcnamara1988use}, and its implement procedure also can be divided into two steps: streaming and collision. As the discrete velocity models are coupled with computational grid, the distribution functions can exact streaming from one grid node to its neighbour in one time step. Such couple result in only the regular grid (Cartesian grid) can be used in standard LBM. Though the mesh generation is easy, using Cartesian grid to simulate relatively high Reynolds number flows is unwise as huge amount grid nodes have to be used to resolve boundary layer flow. As the standard LBM is a special finite-difference scheme of the continuous Boltzmann equation\cite{he1997theory}, the mature finite volume method (FVM) and corresponding numerical schemes developed in macro method also can be transplant into the LBM framework. In the finite volume LBM (FVLBM), the body-fitted grid and hybrid grid can be used, the amount of grid cells can much decline compared with standard LBM, so the FVLBM complete remove the defect of standard LBM on the grid problem. Detailed introductions and discusses about FVLBM can be found in the first part of this paper.

For the turbulent flow simulation, in the macro method, as the resolved scales is different, it can be classified into three methods: direct numerical simulation (DNS), large eddy simulation (LES), and Reynolds-averaged Navier-Stokes equation method (RANS). For DNS, which need not any turbulence model, as there exist a wide range of eddy scales, to simulate the evolution of the smallest eddy, the total amount of grid cells is astronomical and the simulate time is intolerable, so usually the relatively low Reynolds number flows can be simulated at present. With the enhancement of computational power, people develop the LES method, which to imitate the effect of turbulence with sub-grid model. In comparison with DNS, the LES can much decline the amount of grid cells, and turbulent statistical data also can be obtained. Nowadays, the computational cost with DNS and LES for relatively high Reynolds number flow simulation is unacceptable, by contrast, the RANS is a most cost-effective way for engineering applications. Like macro method, the LBM also can be classified as three methods: LBM-DNS, LBM-LES, and LBM-RANS approach\cite{guo2013lattice}. A brief review can be found in Ref. \cite{jahanshaloo2013review}. For LBM-DNS and LBM-LES, same as to macro method, huge amount of grid cells and long computing time result in the relatively low Reynolds number flows can be simulated and limit its range of application. For DNS-LBM, the represent work can be found in Ref. \cite{chikatamarla2010lattice}. For LES-LBM, Hou et al.\cite{hou1994lattice} firstly introduced the subgrid model into the LBM framework, and high Rynolds number cavity flow are simulated to verify the performance the subgrid model. To improve the computational efficiency, Yao et al.\cite{yao2017adaptive} and Guo et al.\cite{guo2015hybrid} developed an adaptive-gridding lattice Boltzmann method, and coulped the LES subgrid model to simulate the flow over the blocks at relative higher Reynolds numbers. Zhuo et al.\cite{zhuo2013based, zhuo2016based} studied the performance of a new LBM model, that is filter-matrix lattice Boltzmann model, in turbulence flow simulations with LES method. Besides, Li et al.\cite{li2014parallel} also developed a new LBM-LES framework on multi-GPUs to improve the computational efficiency. For the LBM-RANS, the implemented method is that the turbulence models developed in macro method are used to calculate the turbulent eddy viscosity and the effective relaxation time is modified by total viscosity (equal to add the kinematic viscosity of fluid to the turbulent eddy viscosity) to model the effect of turbulence\cite{chen2003extended,teixeira1998incorporating}. In standard LBM, $k-\epsilon$, RNG $k-\epsilon$ and  Spalart-Allmaras (S-A) turbulence model have been introduced into its framework to simulate relatively high Reynolds number and obtain good results through test case of flow in pipe and flow around backward-facing step and airfoils\cite{teixeira1998incorporating,filippova2001multiscale,pellerin2015implementation}. Besides, though the multi-blocks and grid refinement techniques can be used, the amount of grid cells is still larger\cite{pellerin2015implementation} than the macro method's. In non-standard LBM, Shu et al.\cite{shu2006application} used Taylor-series-expansion and least-squares-based LBM coupled with $k-\omega$ and Spalart-Allmaras (S-A) turbulence model to simulate the backward-facing step flow at Reynolds number 44000 and obtain good results compared with experimental data, and only about 21 thousand grid cells are used, at same Reynolds number, the amount of grid cells in standard LBM will be much larger. Imamura et al.\cite{imamura2004flow} used generalized form of interpolation-based LBM coupled with Baldwin-Lomax turbulence model to simulate the flow around NACA0012 airfoil at Reynolds number $5\times10^5$ with the amount of grid cells is about 52 thousand. Similarly, with the implementation of interpolation-based LBM on the non-body-fitted non-uniform Cartesian meshes, and coupled with S-A turbulence model, Li et al.\cite{li2012non} also got good results for turbulent flow over the NACA0012 airfoil and the inclined flat plate. The amount of grid also declined compared with that used in standard LBM. So, in the non-standard LBM, usually the amount of grid cells can be declined to the level of macro method's. For the FVLBM-RANS, several pioneering works based on structured grid have been conducted. Choi et al.\cite{choi2010simple} have coupled the $k-\epsilon-\overline{\nu^2}$ turbulence model into FVLBM and have good results for the case of turbulent backward-facing step flow at Reynolds number 5000. Guzel et al.\cite{guzel2015simulation} have coupled the S-A turbulence model to simulate the turbulent flow over the flat plate and the NACA0012 airfoil at much higher Reynolds number. Good results are achieved for both test cases. The studies of FVLBM coupled with others turbulence models are scanty and need further study. To the best of our knowledge, though the $k-\omega$ SST turbulence model is a most popular turbulence model in fluid mechanics, it has not been introduced into the LBM framework.

In non-standard LBM, though the disadvantage about the grid problem in standard LBM is removed, new trouble is emerged. For relatively high Reynolds number flows, the relaxation time decided by kinematic viscosity of fluid is very small, if the collision term is treated with explicit method, the computational efficiency is much lower\cite{rossi2005unstructured}. To remove this defect, the implicit-explicit (IMEX) Runge--Kutta temporal discretization method is introduced into finite difference LBM by Wang et al.\cite{wang2007implicit}. In the implement, the advective term is treated by explicit and collision term is treated by implicit method, based on the property of the collision invariants of the LBM, the implicitness can be eliminated completely. The IMEX method maintains the simplicity of original LBM and the computational efficiency can be much enhanced. This idea now has transplanted into FVLBM framework based on structured grid. Taking advantages of IMEX scheme, Guzel et al.\cite{guzel2015simulation} simulated the turbulent flow at million order of Reynolds number, if explicit Euler method is used, the computing time is unimaginative. Li et al.\cite{li2016aeroacoustic} has used IMEX to simulate compressible flows around NACA0012 airfoil with FVLBM and the numerical stability is also enhanced.

The remainder of this paper is organized as flows. Section \ref{turbulencemodel} presents the turbulence model used in this study. The coupled method between turbulence model and FVLBM scheme, and the acceleration method, namely IMEX scheme are presented in Section \ref{FVLBM}. In Section \ref{Numerical experiments} several test cases are conducted to validate the method proposed in this paper. Finally. Section \ref{Conclusions} is the summary of present method.

\section{Turbulence model}\label{turbulencemodel}
In this paper, the standard $k-\omega$ SST turbulent model\cite{menter1994two} is used for modelling the effect of turbulence. The start point of this model is combine the advantages of $k-\epsilon$ and $k-\omega$ turbulence models for simulating the adverse pressure gradients and separation in aerodynamics and has extended to many other fields since then\cite{menter2003ten}. The brief descriptions of this turbulence model are presented in this section.

The transport equations for the turbulent kinetic energy $k$ and the special dissipation $\omega$ are given by

\begin{equation}\label{SSTequ}
	\frac{\partial{\bm{W}}}{\partial{t}} + \nabla (\bm{F_c}- \bm{F_v}) = \bm{Q},
\end{equation}
where $\bm{W}=(\rho{k},\rho\omega)^T$, $\bm{F_c}$, $\bm{F_v}$ and $\bm{Q}$ are the vectors of convection, diffusion and source terms, respectively, and are given as

\begin{equation}
    \bm{F_c}=(\rho{k}\bm{U}, \rho\omega\bm{U})^T,
\end{equation}

\begin{equation}
    \bm{F_v}=(-(\mu + \sigma_k\mu_t)\nabla k, -(\mu + \sigma_{\omega}\mu_t)\nabla \omega)^T,
\end{equation}

\begin{equation}
    \bm{Q}=(P-\beta^*\rho \omega k, \frac{\rho \gamma P}{\mu_t}-\beta^*\rho \omega^2 + 2(1-f_1)\frac{\rho \sigma_{\omega 2}}{\omega}\nabla k \nabla \omega)^T,
\end{equation}
where $\rho$ and $u$ are the density and velocity of fluid, respectively, $\mu$ and $\mu_t$ are the dynamic molecular viscosity and turbulent eddy viscosity, respectively. $P$ is the production of kinetic energy and is given by

\begin{equation}
       P=\tau_{ij}S_{ij},
\end{equation}
where
\begin{equation}
       \tau_{ij}=\mu_t(2S_{ij}-\frac{2}{3}\frac{\partial{u_k}}{\partial{x_k}}\delta_{ij})-\frac{2}{3}\rho{k}\delta_{ij}, S_{ij}=\frac{1}{2}(\frac{\partial{u_i}}{\partial{x_j}}+\frac{\partial{u_j}}{\partial{x_i}}).
\end{equation}
$f_1$ is the blending function and is given as

\begin{equation}
 f_1 = tanh(arg_1^4), arg_1=\min\left[\max \left(\frac{\sqrt{k}}{0.09\omega d},\frac{500\mu}{\rho \omega d^2} \right), \frac{4\rho\sigma_{\omega2}k}{CD_{k\omega}d^2}  \right],
\end{equation}
where $CD_{k\omega}$ is defined as

\begin{equation}
	CD_{k\omega}=\max \left(\frac{2\rho \sigma_{\omega 2}}{\omega}\frac{\partial k}{\partial x_j}\frac{\partial \omega}{\partial x_j}, 10^{-20} \right),
\end{equation}
and $d$ is the distant to the nearest wall.

The macro physical variables calculated from the FVLBM scheme presented in the first part of this paper will used to solve the Eq.~(\ref{SSTequ}) and the turbulent eddy viscosity can be calculated as

\begin{equation}
	\mu_t = \frac{a_1\rho k}{\max(a_1\omega ,Sf_2)},
\end{equation}
where $S=\sqrt{2S_{ij}S_{ij}}$, and

\begin{equation}
	f_2 = tanh(arg_2^2), arg_2=\max \left(\frac{2\sqrt{k}}{0.09\omega d}, \frac{500\mu}{\rho \omega d^2} \right).
\end{equation}
All the constants presented above are defined as
\begin{equation}\nonumber
\begin{aligned}
	a_1 &= 0.31,\quad \beta^* = 0.09,\quad \kappa = 0.41, \\
	\sigma_{k1} &= 0.85,\quad \sigma_{\omega1} = 0.5,\quad \beta_1 = 0.075, \\
	\sigma_{k2} &= 1.0,\quad \sigma_{\omega2} = 0.856,\quad \beta_2 = 0.0828,\\
   C_{\omega 1} &= \beta_1/\beta^{\ast}-\sigma_{\omega 1}\kappa^2/\sqrt{\beta^{\ast}}, C_{\omega 2} = \beta_2/\beta^{\ast}-\sigma_{\omega 2}\kappa^2/\sqrt{\beta^{\ast}}.
\end{aligned}
\end{equation}

\section{Finite volume lattice Boltzmann method solution procedure for turbulent flow simulation}\label{FVLBM}

\subsection{The improved FVLBM scheme}
The main problem of original FVLBM scheme present in the first part of this paper is poor computational efficiency for turbulent flows. To alleviate this defect and retain the simpleness of  numerical method, implicit-explicit (IMEX) Runge-Kutta temporal discretization scheme is introduced into present FVLBM scheme. Eq.~(\ref{bgkeq}) is the discretized Boltzmann equations,

\begin{equation}\label{bgkeq}
     \frac{\partial{f_{\alpha}(\bm{x},t)}}{\partial{t}} + \bm{e}_{\alpha} \cdot \nabla f_{\alpha}(\bm{x},t)
  = -\frac{1}{\tau}[f_{\alpha}(\bm{x},t) - f_{\alpha}^{eq}(\bm{x},t)],
\end{equation}
where the definitions of $f_\alpha$, $f_{\alpha}^{eq}$, $\bm{e}_{\alpha}$ and $\tau$ can be found in the first part of this paper. The D2Q9 lattice model\cite{qian1992lattice} is also used in this part of paper. Eq.~(\ref{bgkeq}) is typical hyperbolic systems with relaxation and $\tau$ is the stiffness parameter\cite{pareschi2005implicit}. The stability condition result in the time step must less than $2\tau$ when the collision term is treated with explicit method, then this restriction will lead to much long time are needed to simulate the turbulent flows. To improve the computational efficiency, the implicit method must be used. As a consequence, using the IMEX scheme to discrete the Eq.~(\ref{bgkeq}), the convection term is treated with explicit method and collision term is treated with implicit method.

To solve Eq.~(\ref{bgkeq}) with finite volume method and with semi-discrete scheme, it can be rewriten as
\begin{equation}\label{semibgk}
      \frac{d}{dt}f_\alpha = \Gamma^a_\alpha+\Gamma^c_\alpha,
\end{equation}
where $\Gamma^a_\alpha$ is the convection term and is given by
\begin{equation}\label{advection}
      \Gamma^a_\alpha = -\frac{1}{A}\sum_{m}\left[(\bm{e}_\alpha \cdot \bm{n})f_{\alpha,bc}\Delta{l}\right]_m,
\end{equation}
$\Gamma^c_\alpha$ is the collision term and is given by
\begin{equation}\label{collision}
      \Gamma^c_\alpha = -\frac{1}{\tau}\left[f_{\alpha}(\bm{x},t) -f_{\alpha}^{eq}(\bm{x},t)\right].
\end{equation}
The definitions of symbol in Eq.~(\ref{advection}) and Eq.~(\ref{collision}), and corresponding calculation methods also can be found in the first part of this paper.

In our work, three-stage, second-order IMEX scheme is used to discrete the Eq.~(\ref{semibgk}) and given by
\begin{subequations}\label{imexshceme}
\begin{equation}\label{imexone}
      f_\alpha^{(J)} = f_\alpha^n + \Delta{t}\sum_{k=1}^{J-1}\widetilde{m}_{Jk}\Gamma^{a,(k)}_\alpha +                   \Delta{t}\sum_{k=1}^{J}m_{Jk}\Gamma^{c,(k)}_\alpha,
\end{equation}
\begin{equation}\label{imextwo}
      f_\alpha^{n+1} = f_\alpha^n + \Delta{t}\sum_{J=1}^{3}\widetilde{n}_{J}\Gamma^{a,(J)}_\alpha +                   \Delta{t}\sum_{J=1}^{3}n_{J}\Gamma^{c,(J)}_\alpha,
\end{equation}
\end{subequations}
where $n+1$ and $n$ represent the distribution function at two time levels, respectively, and $J$ represents the stage number. $\widetilde{m}_{Jk}$ and $m_{Jk}$ are the $3\times3$ metrics, $\widetilde{n}_{J}$ and $n_{J}$ are the $3$-dimensional vectors, and these characteristic coefficients are given by

\begin{equation}
 \begin{aligned}
 & \widetilde{m}_{Jk}  =  \left(
                          \begin{array}{ccc}
   	                      	0    &  0   &   0  \\
   	    	                1/2  &  0   &   0  \\
   	                    	1/2  &  1/2 &   0  \\
                        \end{array}
                        \right),
             m_{Jk}  =  \left(
                       \begin{array}{ccc}
	                     	1/4  &  0   &   0    \\
    	                    0    &  1/4 &   0    \\
    	                 	1/3  &  1/3 &   1/3  \\
                       \end{array}
                       \right), \\
 & \widetilde{n}_{J}  = \left(
                         \begin{array}{ccc}
                            1/3  &  1/3 &   1/3
                         \end{array}
                       \right),
              n_{J}  = \left(
                         \begin{array}{ccc}
                            1/3  &  1/3 &   1/3
                         \end{array}
                       \right). \\
   \end{aligned}
\end{equation}

The difficulty to solve the Eq.~(\ref{imexshceme}) is that at the stage $J$, both $f_\alpha^{(J)}$ and $f_\alpha^{eq,(J)}$ in $\Gamma_\alpha^{c, (J)}$ are unknown and traditional implicit methods are not easy to implement as $f_\alpha^{eq,(J)}$ and $f_\alpha^{(J)}$ all are needed to be fixed in iteration\cite{wang2007implicit}. Thanks to the characteristic of collision invariants of LBM, an ingenious method can be used to deal with this problem. Eq.~(\ref{imexone}) can be rewrite as

\begin{equation}\label{sumimeximplicit}
      \sum_{\alpha}f_\alpha^{(J)}\phi = \sum_{\alpha}f_\alpha^n\phi + \Delta{t}\sum_{k=1}^{J-1}\widetilde{m}_{Jk}\left[\sum_{\alpha}\Gamma^{a,(k)}_\alpha\phi\right] +                   \Delta{t}\sum_{k=1}^{J}m_{Jk}\left[\sum_{\alpha}\Gamma^{c,(k)}_\alpha\phi\right],
\end{equation}
and the characteristic of collision invariants can be defined as

\begin{equation}\label{collisionvariant}
          \sum_{\alpha}(f_\alpha^{eq,(k)}-f_\alpha^{(k)})\phi=0,
\end{equation}
where $\phi=(\rho,u,v)$ is the vector of macro physical variables. Substituting Eq.~(\ref{collisionvariant}) into Eq.~(\ref{sumimeximplicit}) results in

\begin{equation}\label{sumimeximplicitnew}
      \sum_{\alpha}f_\alpha^{(J)}\phi = \sum_{\alpha}f_\alpha^n\phi + \Delta{t}\sum_{k=1}^{J-1}\widetilde{m}_{Jk}\left[\sum_{\alpha}\Gamma^{a,(k)}_\alpha\phi\right],
\end{equation}
that is the macro physical variables $\rho^{(J)}$ and $\bm{u}^{(J)}$ can be calculated from the known variables. Then the corresponding $f_\alpha^{eq,(J)}$ also can be evaluated. Finally, the distribution function at stage $J$ can be explicit calculated as

\begin{equation}\label{fnewstage}
	 f_\alpha^{(J)} = \frac{f_\alpha^n + \Delta{t}\sum_{k=1}^{J-1}(\widetilde{m}_{Jk}\Gamma^{a,(k)}_\alpha + m_{Jk}\Gamma_\alpha^{c,(k)}) +     \frac{\Delta{t}}{\tau^{(J)}}m_{JJ}f_\alpha^{eq,(J)}}{1+\frac{\Delta{t}}{\tau^{(J)}}m_{JJ}}.
\end{equation}

Here, a brief comparison of computational cost between the explicit Euler scheme and the IMEX scheme is presented. For the explicit Euler scheme, the convection term and collision term are calculated only once in each time iteration. But, the same calculations will perform three times for the IMEX scheme. In consideration of the computational cost of Eq.~(\ref{imextwo}) and the increase of memory, the IMEX scheme will cost about five times of computational time than the explicit Euler scheme in each time iteration. As the convection term is treated with explicit method, the time step $\Delta{t}$ can not very larger, if $\Delta{t}$ less then $10\tau$ when the IMEX scheme is used, it has any advantages than the explicit Euler scheme. So this improved FVLBM scheme is not suitable for low Reynolds number flow simulations.

\subsection{The coupling scheme for turbulent flow}
In this paper, the finite volume method also used to solve the $k-\omega$ SST turbulence model. Eq.(\eqref{SSTequint}) is the integral form of $k-\omega$ SST model:
\begin{equation}\label{SSTequint}
     \frac{\partial}{\partial{t}}{\int_\Omega \rho\bm{U} d\Omega}
     + {\oint_{\partial\Omega} (\bm{F}_c - \bm{F}_v) \cdot \bm{n} dl} =
     {\int_{\Omega} \bm{Q} d\Omega},
\end{equation}
where $\bm{U}$ represent the turbulent kinetic energy $k$ and the special dissipation $\omega$, respectively. The definitions and expressions of $\bm{F}_c$, $\bm{F}_v$ and $\bm{Q}$ are illustrated in Eq.~(\ref{SSTequ}). The same definitions used in FVLBM scheme also can illustrate the other symbols in Eq.~(\ref{SSTequint}).

The same grid systems used for FVLBM are also used for discretization form of the Eq.~(\ref{SSTequint}). The second-order upwind scheme is used to discretize the $\bm{F}_c$ and center difference scheme is used to discretize the $\bm{F}_v$. The temporal discretization scheme is an implicit method presented in Ref. \cite{archambeau2004code}. The boundary condition presented in Ref. \cite{menter1994two} will also used here.

If turbulent kinematic viscosity $\nu_t$ is calculated, it will used to modify the relaxation time $\tau$ in Eq.~(\ref{collision}) as
\begin{equation}\label{turbtau}
          \tau=\nu_{total}/c_s^2=(\nu + \nu_t)/c_s^2,
\end{equation}
where $\nu$ is the kinematic viscosity of fluid and $c_s^2$ is the sound speed of lattice (equal to $1/3$ for D2Q9 lattice model).

For the coupling of FVLBM and SST model, in each iteration, once the FVLBM completes its calculation, the updated macro physical variables will be used to solve the SST model, then Eq.~(\ref{turbtau}) will be calculated and used to modify the relaxation time in Eq.~(\ref{collision}) for next calculation.

At the end of this section, the general implementation of present improved FVLBM is detailed as follows:
\begin{description}
\item[Step 0.] Initialize $\rho$, $\bm{u}$ on the computational domain, use their values to initialize the $f_{\alpha}^{eq}$ and set $f_{\alpha}=f_{\alpha}^{eq}$. Initialize $k$, $\omega$ if $k-\omega$ SST turbulence model is activated.
\item[Step 1.] Solve the $k-\omega$ SST turbulence model and use Eq.~(\ref{turbtau}) to modify the relaxation time if turbulence model is activated.
\item[Step 2.] Compute the gradient of $f_\alpha$ in each cell and boundary conditions with method presented in the first part of this paper. Use Eq.~(\ref{advection}) to evaluate the convection term.
\item[Step 3.] Use Eq.~(\ref{sumimeximplicitnew}) to calculate the macro physical variables at new stage, then evaluate the $f_\alpha^{eq,(J)}$.
\item[Step 4.] Use Eq.~(\ref{collision}) to evaluate the collision term and use Eq.~(\ref{fnewstage}) to calculate the distribution function $f_\alpha$ at new stage.
\item[Step 5.] Repeat Steps 1-4 to calculate the $f_\alpha$ at other two stages.
\item[Step 6.] Update the $f_\alpha$ from $f_\alpha^n$ to $f_\alpha^{n+1}$ according to Eq.~(\ref{imextwo}).
\item[Step 7.] Update the macro physical variables $\rho$, $p$ and $\bm{u}$ in each cell.
\item[Step 8.] Go back to Step 1 to start a new iteration.
\end{description}

\section{Numerical experiments}\label{Numerical experiments}
In this section, three flow problems are simulated to validate the method used in this paper. The first case is lid-driven square cavity flow, which used to test the acceleration performance of IMEX scheme. Then, two turbulent flow cases, namely flow over the backward-facing step and flow around the NACA0012 airfoil, are simulated to validate the code developed in this paper. The unstructured mesh used in this section are also generated through Salome, an open-source software that provides a generic platform for Pre- and Post-Processing for numerical simulations (\url{http://www.salome-platform.org/}).

Before the discussions of numerical result, the framework of code is outlined briefly again. The  present FVLBM scheme coupled with $k-\omega$ SST turbulence model also has been coded with the help of Code\underline{ }Saturne\cite{archambeau2004code}, an open-source CFD software of EDF, France (\url{http://code-saturne.org/cms/}). In our work, the original codes of $k-\omega$ SST turbulence model in Code\underline{ }Saturne are modified a little and coupled with the original codes of FVLBM. Besides, the new developed codes of FVLBM also have the ability of parallel computing.

\subsection{Lid-driven square cavity flow}

\begin{figure}
 	\centering
 	\subfigure{
 			\includegraphics[width=0.45 \textwidth]{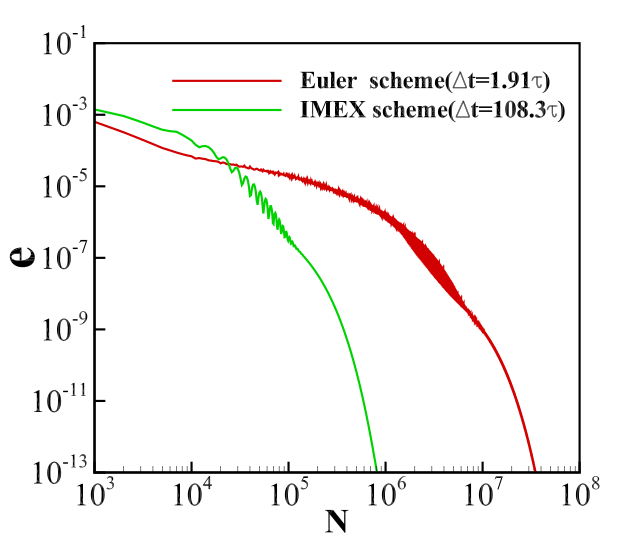}
 			}
 	\caption{\label{IMEXconvergence} Time histories of simulation by explicit Euler scheme and IMEX scheme for the lid-driven square cavity flow at $Re = 3200$.}
\end{figure}

\begin{figure}
 	\centering
 	\subfigure[]{
 			\includegraphics[width=0.45 \textwidth]{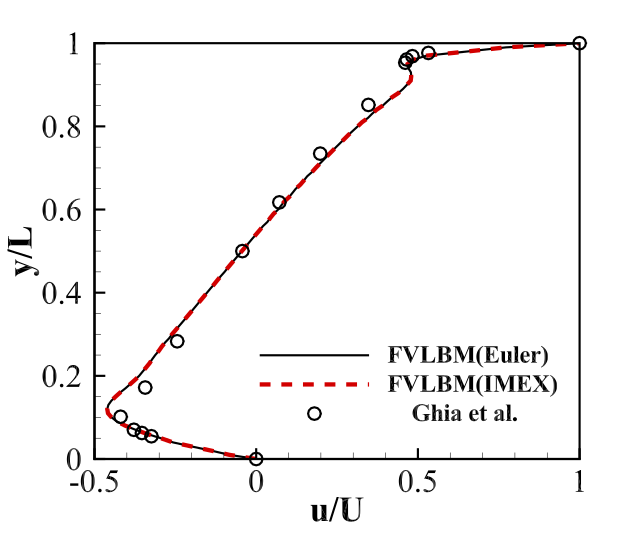}
 			}
    \subfigure[]{
     		\includegraphics[width=0.45 \textwidth]{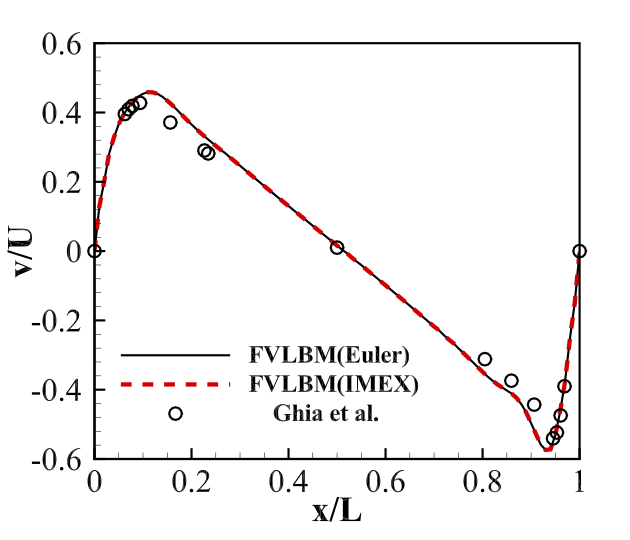}
         	}
 	\caption{\label{cavityIMEX} The velocity profiles at (a) the horizontal and (b) the vertical centerline of cavity at $Re = 3200$\cite{ghia1982high}.}
\end{figure}

To evaluate the acceleration performance of IMEX temporal discretization scheme, the lid-driven square cavity flow is simulated again. For the grid, the total of $128^2$ right triangles which used to grid-convergence studies in the first part of this work is chose here. The Reynolds number is equal to $3200$. The initial conditions and boundary conditions are also same as to that case. The time step is set to $1.91\tau$ for explicit Euler scheme and set to $108.3\tau$ for IMEX scheme, both time steps are the largest value allowed by stability criterion for the corresponding temporal discretization formulations. Fig.~\ref{IMEXconvergence} shows the convergence history of velocity residual $e$ at every 1000 iteration steps $N$. When $e$ reach $1.0\times10^{-13}$, the IMEX scheme need about 800 thousand iterations and the explicit Euler scheme need about 35 million iterations, so the simulating time may decline to $1/9$ of original FVLBM scheme (here we consider that the computational cost of IMEX scheme is about $5$ times greater than that for the explicit Euler schemes for each iteration). Besides, we have compared the velocity profiles resulted from present two FVLBM schemes at the horizontal and  vertical centerline of cavity with results obtained by Ghia et al. (see Fig.~\ref{cavityIMEX}). It is clear that the difference between two results can be neglected, so the IMEX scheme can accelerate the convergence history about one order of magnitude and good results also can be obtained compared with original explicit Euler scheme.

\subsection{Flow over the backward-facing step}

\begin{figure}
 	\centering
    \subfigure{
     		\includegraphics[width=0.65 \textwidth]{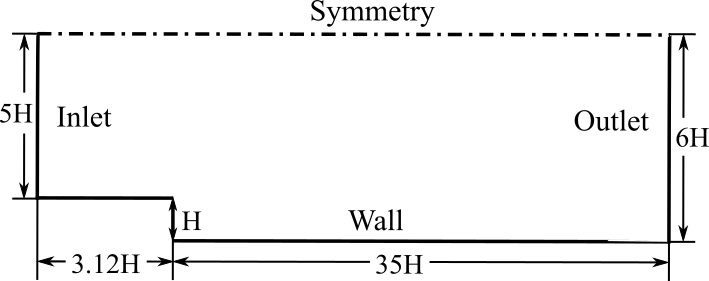}
         	}
 	\caption{\label{backstepconfig} Computational domain for the flow over the backward-facing step (not drawn to scale).}
\end{figure}

\begin{figure}
 	\centering
    \subfigure{
     		\includegraphics[width=0.7 \textwidth]{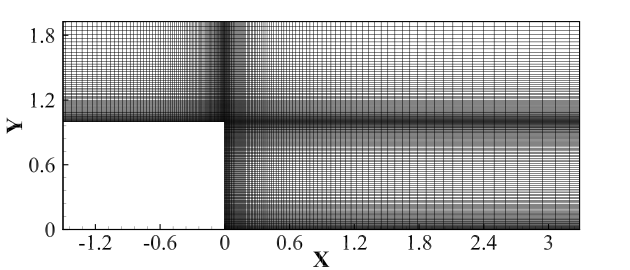}
         	}
 	\caption{\label{backstepgrid} Zoomed view of mesh near the backward-facing step.}
\end{figure}

Flow over the backward-facing step is considered as the first turbulent flow test case in this paper. As its simple configuration, this flow problem is usually taken as a benchmark case to validate the performances of numerical method for predicting the separation flow. Based on the height of step $H$ and the inlet free-stream velocity $U_{ref}$, the flow at Reynolds number $Re = 5000$ is solved. Both experiment and numerical simulations are available in the literatures for this problem\cite{choi2010simple, jovic1994backward,le1997direct,shur1995co}. The size of computational domain and the boundary conditions are shown in Fig.~\ref{backstepconfig}. The grid is used with $41300$ number of quadrangular elements and the part of grid are shown in Fig.~\ref{backstepgrid}. The smallest size of grid near the wall is about $0.004$ and $100$ vertical grid are placed within the step. The $\rho = 1.0$, $u = 0.0$ and $v = 0.0$ are used to initialize the solution domain. For the inlet boundary condition, in our work, the turbulent flow over a flat plate was first simulated to generate proper inlet values that matching the experimental data described in Ref. \cite{jovic1994backward}, then these values are used as inlet boundary condition at location $x = -3.12h$. The IMEX scheme is used as temporal discretization method and set equal to $\Delta{t} = 60\tau$.


\begin{figure}
 	\centering
    \subfigure{
     		\includegraphics[width=0.7 \textwidth]{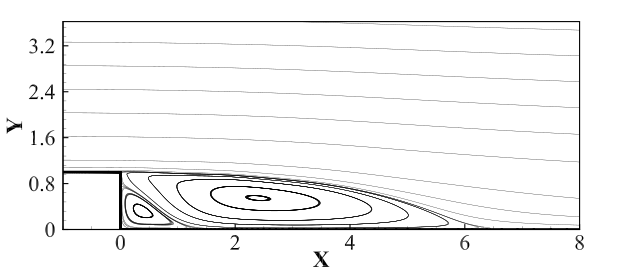}
         	}
 	\caption{\label{bakstepstream} Streamline patterns over the backward-facing step.}
\end{figure}

\begin{figure}
 	\centering
    \subfigure[]{
     		\includegraphics[width=0.45 \textwidth]{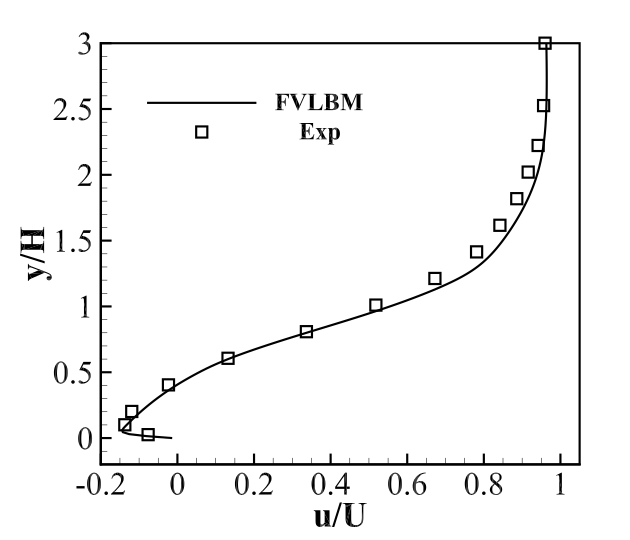}
         	}
    \subfigure[]{
         	\includegraphics[width=0.45 \textwidth]{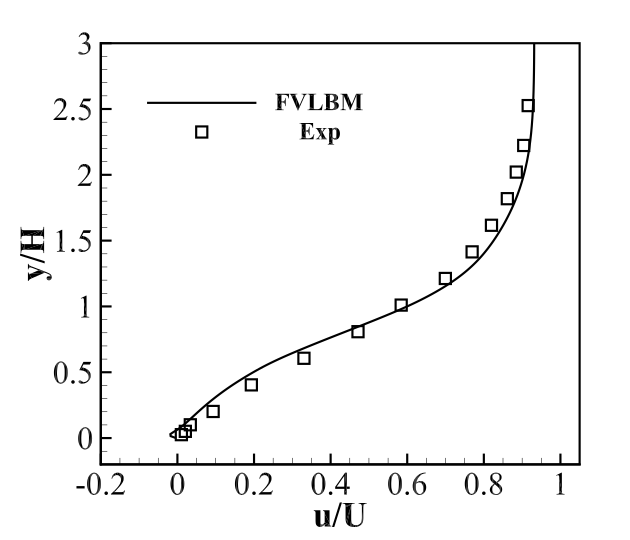}
           	}
    \subfigure[]{
           	\includegraphics[width=0.45 \textwidth]{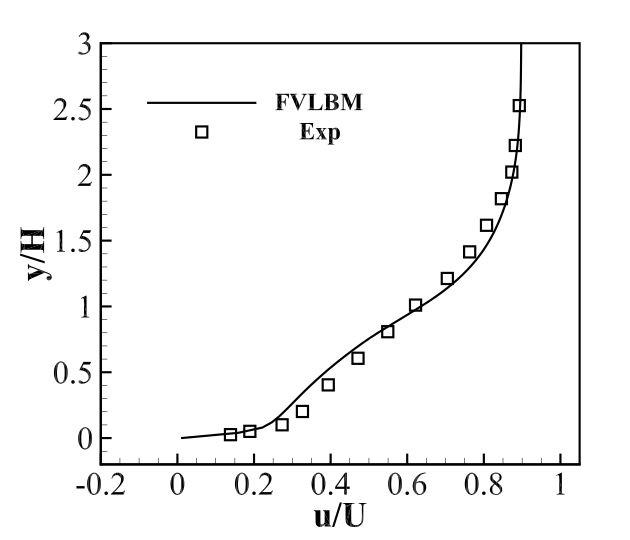}
           	}
    \subfigure[]{
           	\includegraphics[width=0.45 \textwidth]{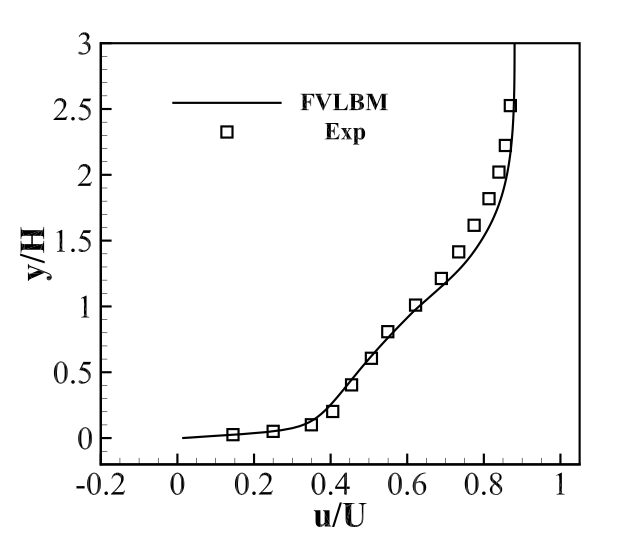}
           	}
    \subfigure[]{
            \includegraphics[width=0.45 \textwidth]{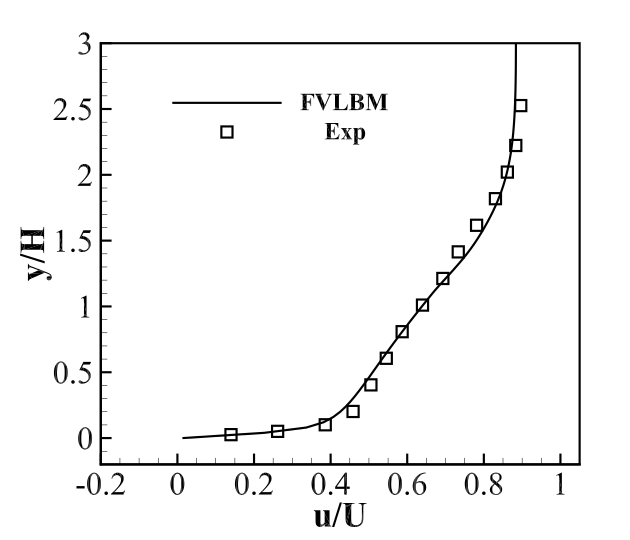}
           	}
 	\caption{\label{bakstepsux} Comparisons of velocity profiles with experimental data obtained by Jovic and Driver\cite{jovic1994backward} at location (a) $x=4h$, (b) $x=6h$, (c) $x=10h$, (d) $x=15h$, and (e) $x=19h$.}
\end{figure}

Fig.~\ref{bakstepstream} shows the streamlines near the step. The predicted length of separation eddy is about $6.5h$ and is close to $6.28h$ obtained by DNS\cite{le1997direct}. Similar to Le et al.'s results\cite{le1997direct}, the secondary eddy at the corner of step is also predicted by present method. Maybe due to the difference of models, this small eddy can not be predicted by Choi et al.\cite{choi2010simple} with $k-\epsilon-\overline{\nu^2}$ turbulence model. Fig.~\ref{bakstepsux} show the comparison of velocity profiles at five locations with experimental data. In general, our results agree well with the experimental data.

\begin{figure}
 	\centering
    \subfigure{
     		\includegraphics[width=0.5 \textwidth]{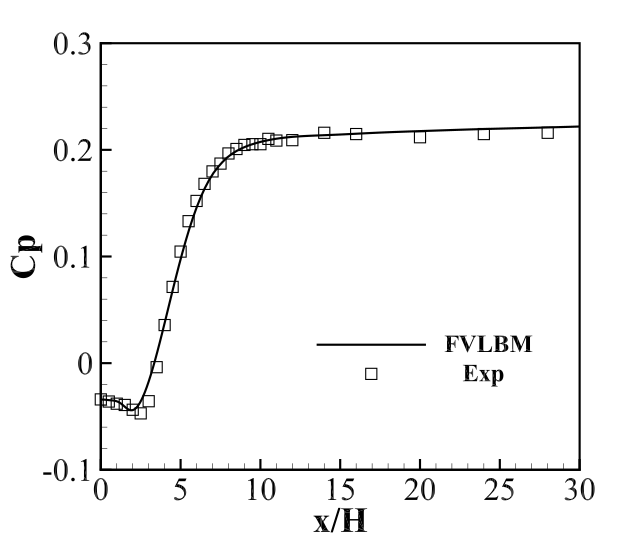}
         	}
 	\caption{\label{backstepCp} Pressure coefficient distribution over the bottom wall of backward-facing step.}
\end{figure}

Fig.~\ref{backstepCp} shows the surface pressure coefficients of bottom wall. It is clear that the present scheme also can obtain good results compared with experimental data. But, in our experiments, due to the inlet and outlet boundary conditions for internal flows presented in the first part of this paper can not balance the total mass in the channel, the oscillation of overall mass is obvious and lead to slower convergence. The technique presented in Ref.\cite{tong2007mass} maybe a possible solution to relieve the oscillation in the computational domain. However, as the inlet and outlet boundary conditions based on the structured grid are not easy to implement on unstructured grid\cite{shu2006application, tong2007mass}, further works will be continued to solve this problem.

\subsection{Flow around the NACA0012 airfoil}

\begin{figure}
 	\centering
 	\subfigure[]{
 			\includegraphics[width=0.45 \textwidth]{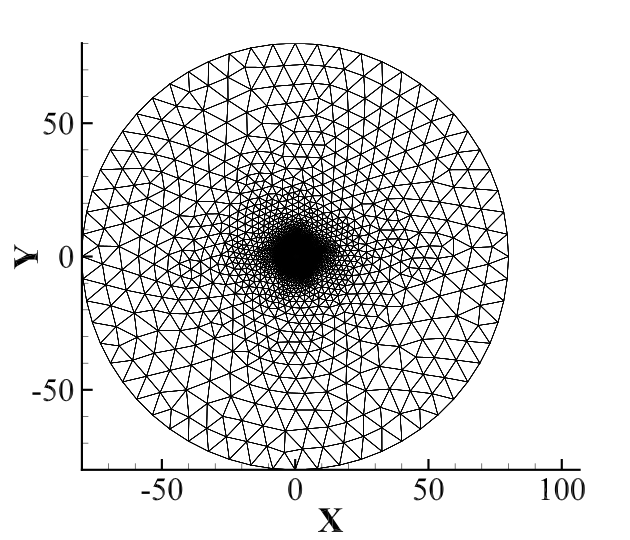}
 			}
    \subfigure[]{
     		\includegraphics[width=0.45 \textwidth]{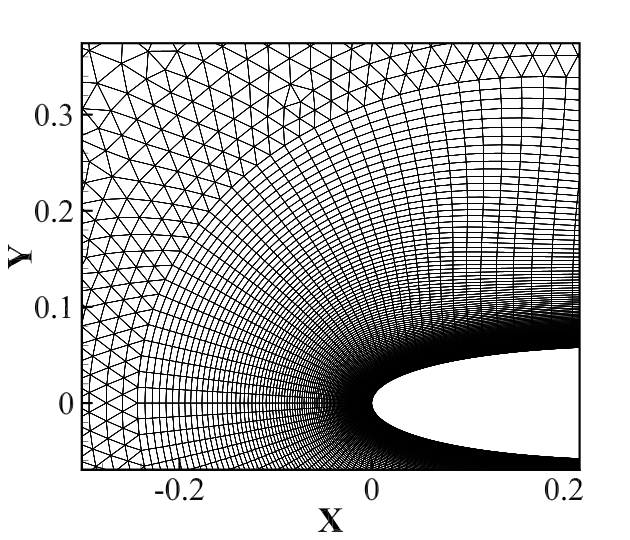}
         	}
 	\caption{\label{naca0012mesh} Hybrid girds for the simulation of turbulent flow around NACA0012 airfoil; (a) Far field and (b) Zoomed view of mesh near the leading edge.}
\end{figure}

The second turbulent flow test case considered in this paper is flow around the NACA0012 airfoil with Reynolds number equal to $5\times10^5$. Here, the Reynolds number Re can be defined as $Re = UL/\nu$, where $U = 0.1$ is the reference velocity, $L = 1.0$ is the chord length of airfoil and $\nu$ is the kinematic viscosity of fluid. Fig.~\ref{naca0012mesh} shows the grid used for this case, 39448 elements (the number of quadrilateral cells is $21600$) are used in the hybrid grid. The computational domain is a circle with radius equal to $80L$. The computational domain is very large so as to eliminate the influence of far-field boundary condition. There are $181$ grid points located at the airfoil and the normal smallest size of grid near the airfoil is $1.0 \times 10^{-4}$. A simple comparison shows that the present FVLBM can much decline the amount of grid cells. For the same case, in macro method, Lockard et al.\cite{lockard2002evaluation} used a C-mesh of $373\times141$ or $52593$ cells with the CFL3D code and the smallest size of grid near the airfoil is $1.2 \times 10^{-4}$. In LBM field, Li et al.\cite{li2012non} used about 4 million grid cells with the generalized form of interpolation supplemented LBM and Pellerin et al.\cite{pellerin2015implementation} used about $550$ thousand grid cells with standard LBM coupled with multi-blocks and grid-refinement techniques. The smallest size of grid in their test cases are $2.0 \times 10^{-4}$ and $1.22 \times 10^{-4}$, respectively. As the Reynolds number equal to $5\times10^5$ is relatively low in aerospace field, to simulate more higher Reynolds number flow such as 6 million turbulent flow, about $10^5$ order of grid are enough to resolve the boundary layer for macro-method\cite{mian2013application} and structured grid FVLBM\cite{guzel2015simulation} and good results can be obtained compared with experimental data, at least to surface pressure coefficients, lift coefficient and drag coefficient. But for standard LBM, even though the multi-blocks and grid-refinement techniques are used, the amount of grid cells will beyond one million at least\cite{pellerin2015implementation}.

In this case, simulations were conducted with the angles of attack $\alpha$ equal to $0^{\circ}$, $3^{\circ}$, $7^{\circ}$ and $12^{\circ}$, respectively. For the initial conditions, the density of fluid is set to $\rho = 1.0$ and the velocities are set to $u = Ucos(\alpha\pi/180)$, $v=Usin(\alpha\pi/180)$. The IMEX temporal discretization scheme is used and time step $\Delta{t} = 180\tau$. Here we demonstrate again that IMEX scheme can much enhance the computational efficiency as the largest time step allowed by explicit Euler scheme is only $\Delta{t} = 2\tau$.

\begin{figure}
 	\centering
 	\subfigure[]{
 			\includegraphics[width=0.45 \textwidth]{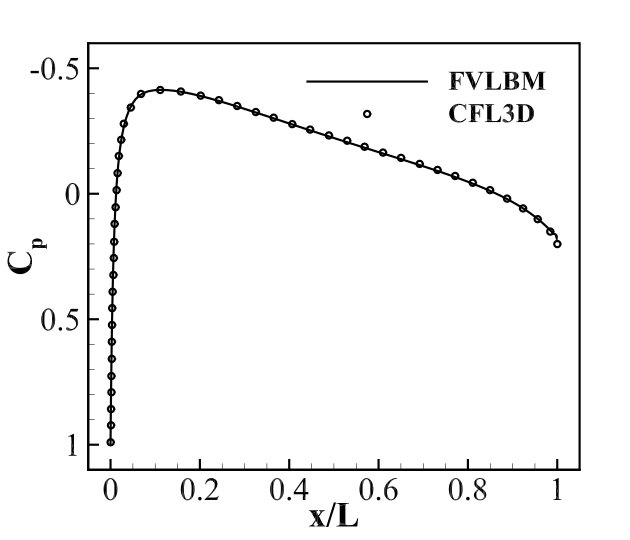}
 			}
    \subfigure[]{
     		\includegraphics[width=0.45 \textwidth]{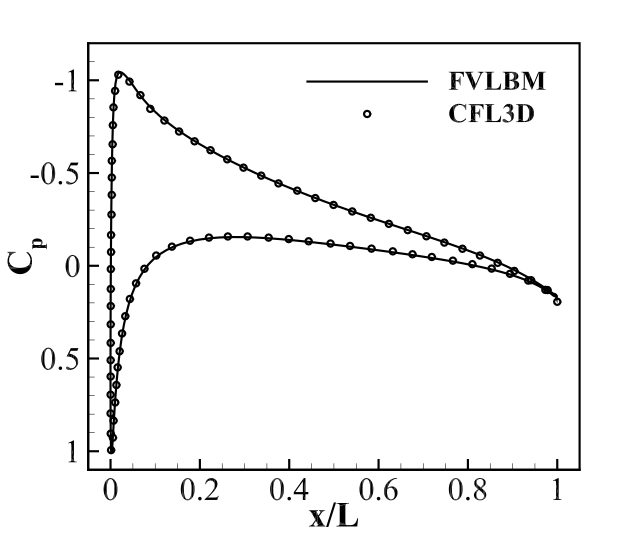}
         	}
    \subfigure[]{
        	\includegraphics[width=0.45 \textwidth]{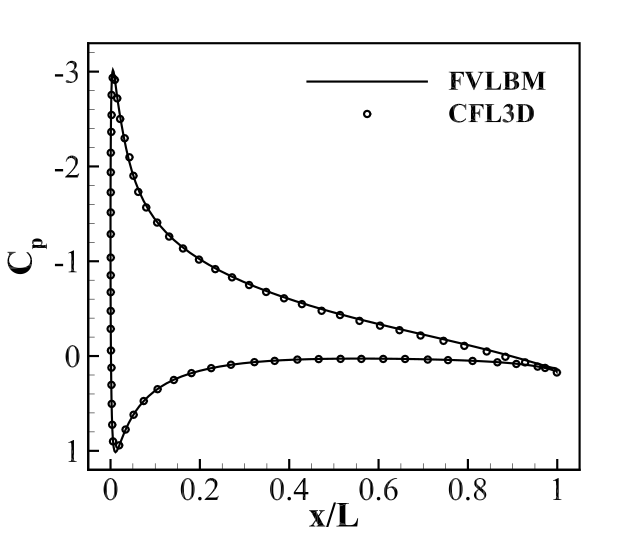}
          	}
    \subfigure[]{
            \includegraphics[width=0.45 \textwidth]{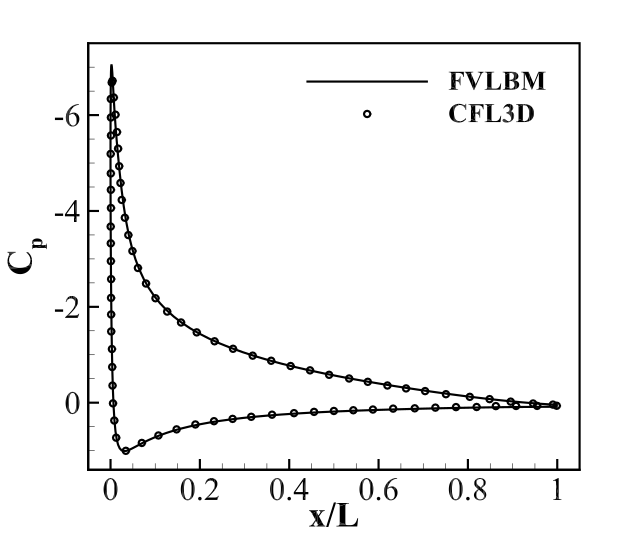}
           	}
 	\caption{\label{naca0012Cp} Pressure coefficient $C_p$ distribution around NACA0012 airfoil at (a) $\alpha = 0^{\circ}$, (b) $\alpha = 3^{\circ}$, (c) $\alpha = 7^{\circ}$, and (d) $\alpha = 12^{\circ}$.}
\end{figure}

\begin{figure}
 	\centering
 	\subfigure[]{
 			\includegraphics[width=0.45 \textwidth]{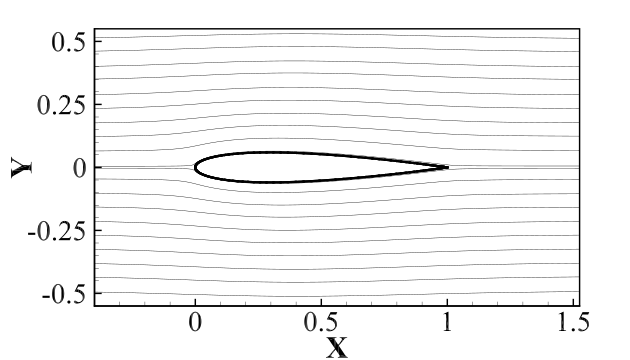}
 			}
    \subfigure[]{
     		\includegraphics[width=0.45 \textwidth]{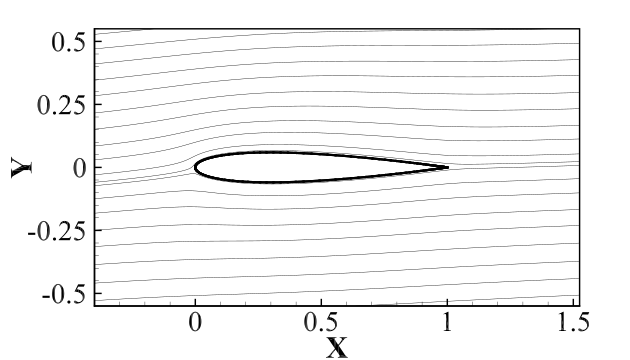}
         	}
    \subfigure[]{
        	\includegraphics[width=0.45 \textwidth]{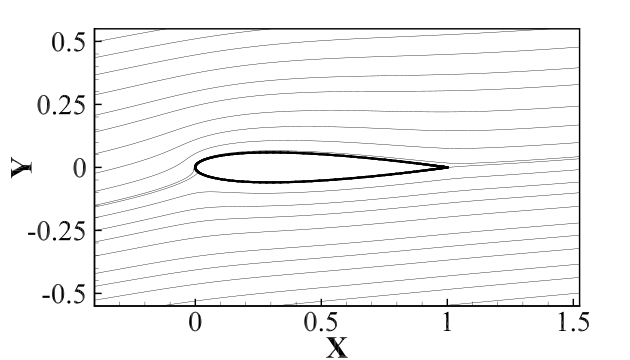}
          	}
    \subfigure[]{
            \includegraphics[width=0.45 \textwidth]{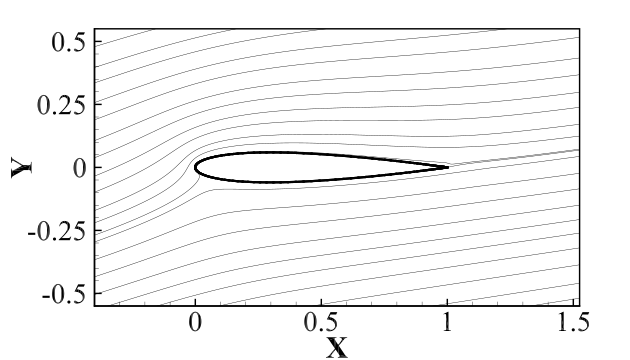}
           	}
 	\caption{\label{naca0012stream} Streamline patterns over NACA0012 airfoil at (a) $\alpha = 0^{\circ}$, (b) $\alpha = 3^{\circ}$, (c) $\alpha = 7^{\circ}$, and (d) $\alpha = 12^{\circ}$.}
\end{figure}

Fig.~\ref{naca0012Cp},\ref{naca0012stream} respectively show the surface pressure coefficients and streamlines around airfoil at different angels of attack (AOA). It is obvious that the present results are in good agreement with the CFL3D data at all four angels of attack. As expected, no flow separation is happened since the separated flow around the airfoil is suppressed when the turbulence model is used.

\begin{figure}
 	\centering
 	\subfigure[]{
 			\includegraphics[width=0.45 \textwidth]{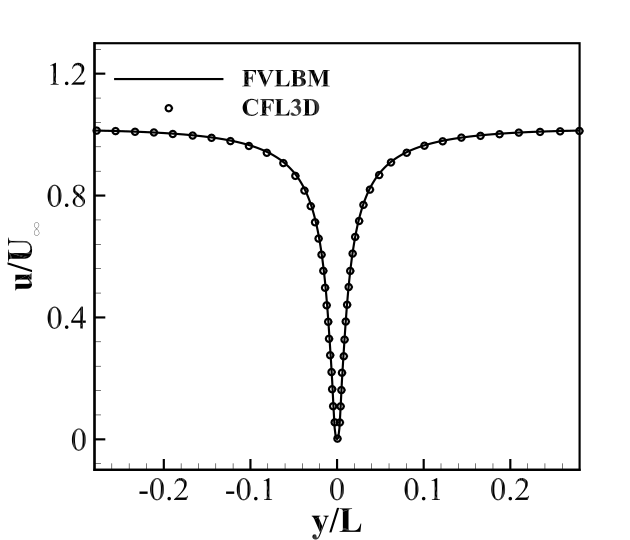}
 			}
    \subfigure[]{
     		\includegraphics[width=0.45 \textwidth]{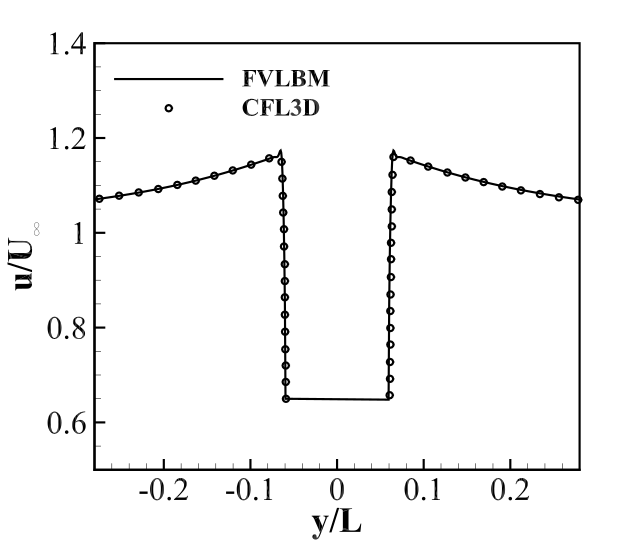}
         	}
    \subfigure[]{
        	\includegraphics[width=0.45 \textwidth]{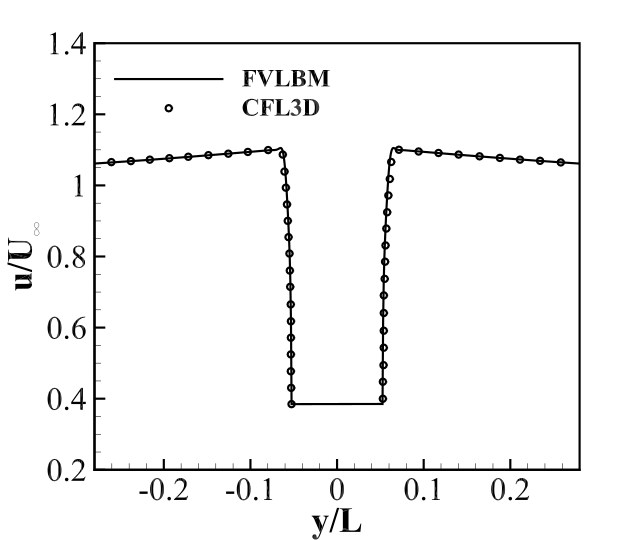}
          	}
    \subfigure[]{
            \includegraphics[width=0.45 \textwidth]{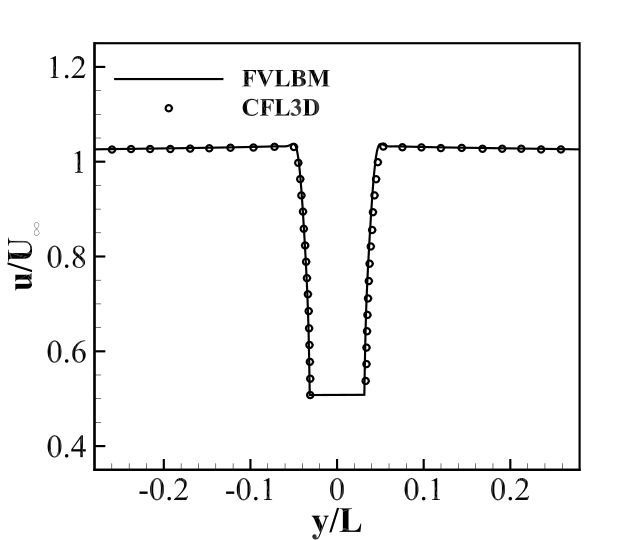}
           	}
    \subfigure[]{
            \includegraphics[width=0.45 \textwidth]{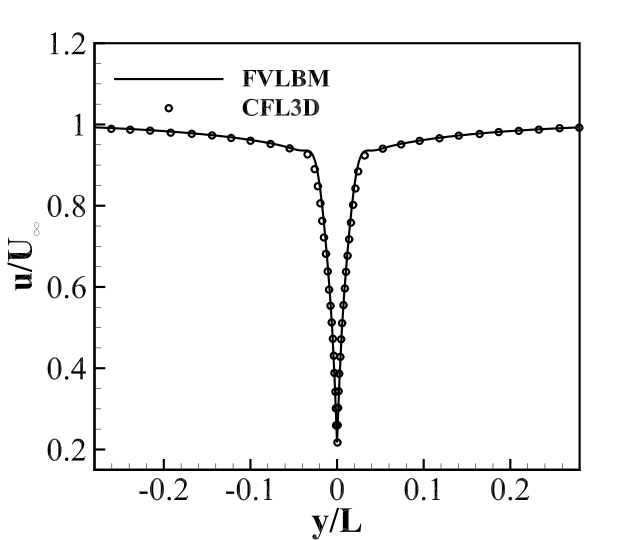}
           	}
 	\caption{\label{naca0012uxalpha0} The x-component of velocity at (a) $x/L = 0.0$, (b) $x/L = 0.25$, (c) $x/L = 0.5$, (d) $x/L = 0.75$, and (e) $x/L = 1$ for flow around NACA0012 airfoil with $\alpha = 0^{\circ}$.}
\end{figure}

\begin{figure}
 	\centering
 	\subfigure[]{
 			\includegraphics[width=0.45 \textwidth]{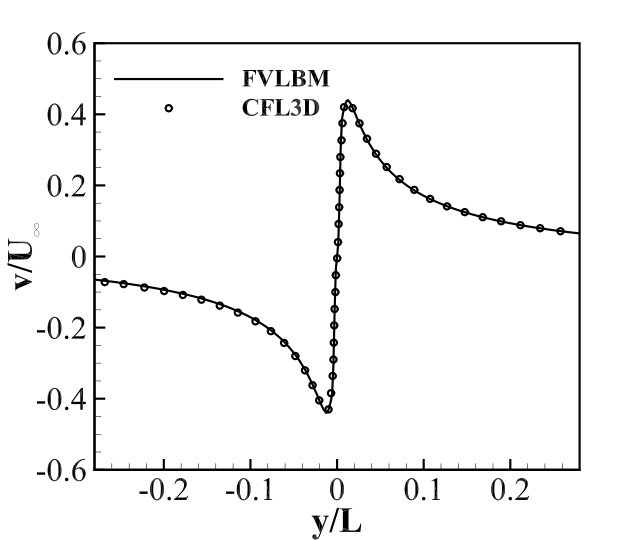}
 			}
    \subfigure[]{
     		\includegraphics[width=0.45 \textwidth]{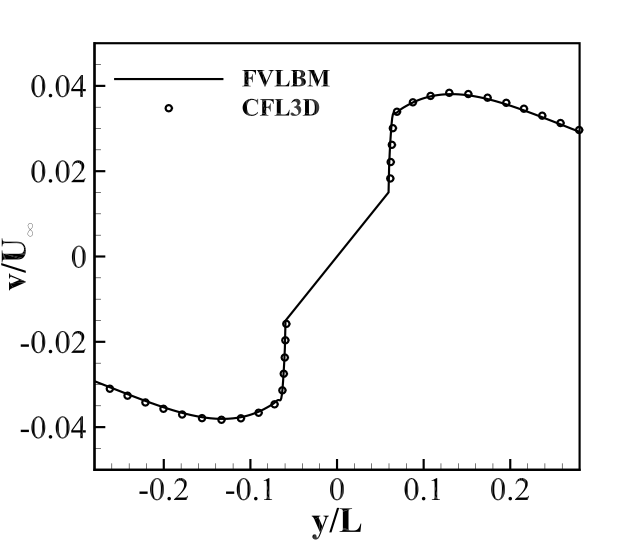}
         	}
    \subfigure[]{
        	\includegraphics[width=0.45 \textwidth]{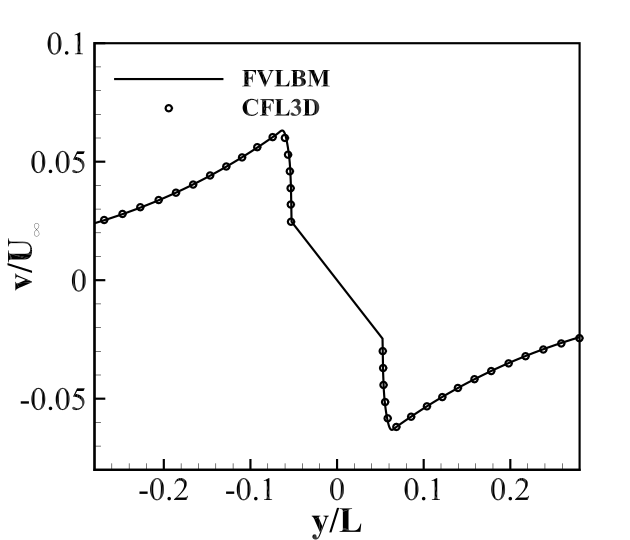}
          	}
    \subfigure[]{
            \includegraphics[width=0.45 \textwidth]{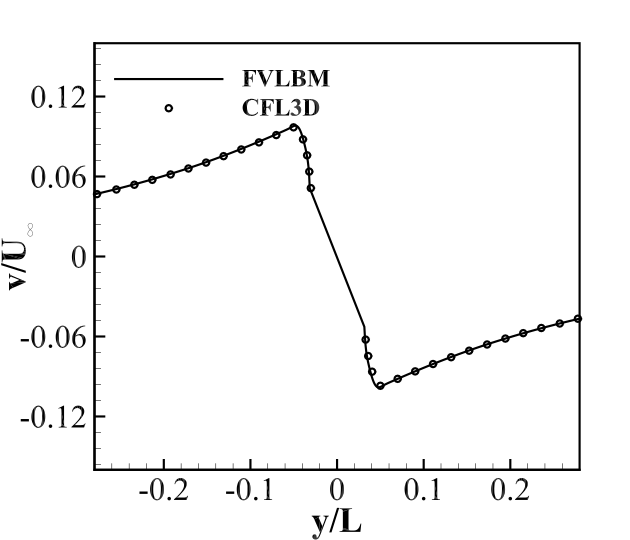}
           	}
    \subfigure[]{
            \includegraphics[width=0.45 \textwidth]{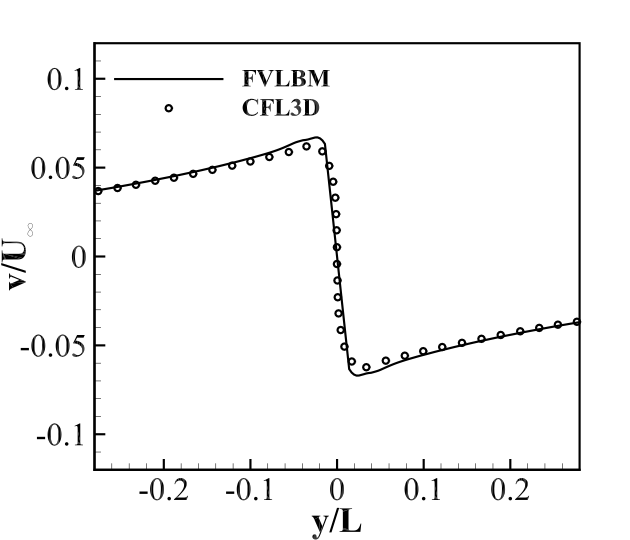}
           	}
 	\caption{\label{naca0012vxalpha0} The y-component of velocity at (a) $x/L = 0.0$, (b) $x/L = 0.25$, (c) $x/L = 0.5$, (d) $x/L = 0.75$ and (e) $x/L = 1$ for flow around NACA0012 airfoil with $\alpha = 0^{\circ}$.}
\end{figure}

\begin{figure}
 	\centering
 	\subfigure[]{
 			\includegraphics[width=0.45 \textwidth]{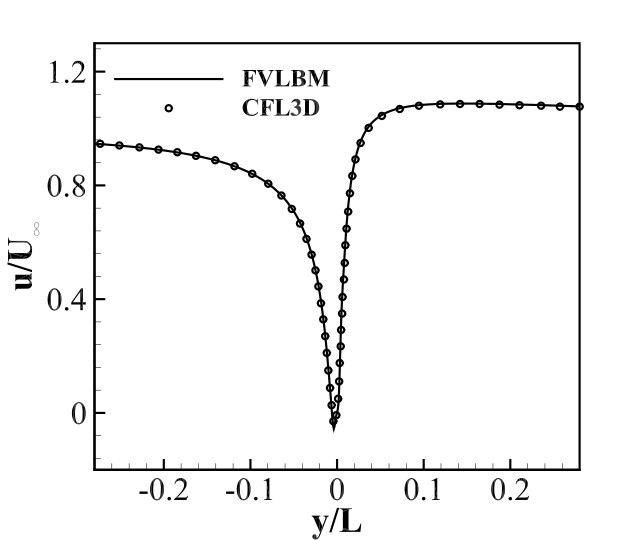}
 			}
    \subfigure[]{
     		\includegraphics[width=0.45 \textwidth]{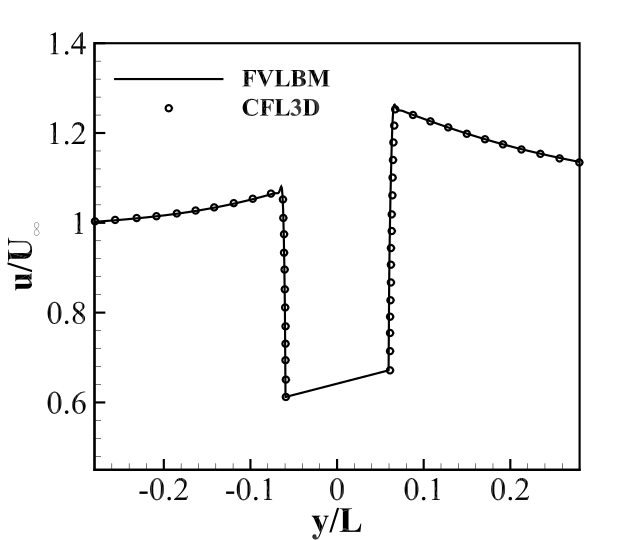}
         	}
    \subfigure[]{
        	\includegraphics[width=0.45 \textwidth]{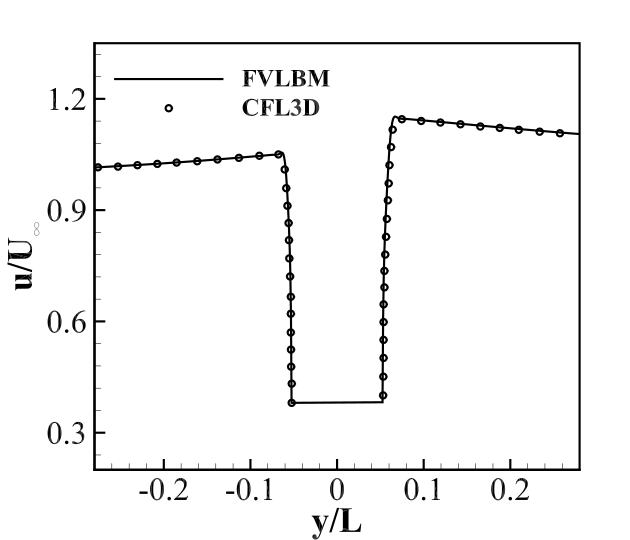}
          	}
    \subfigure[]{
            \includegraphics[width=0.45 \textwidth]{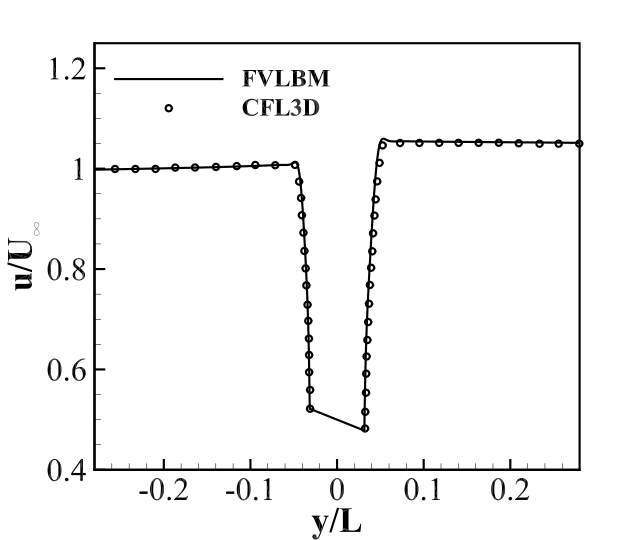}
           	}
    \subfigure[]{
            \includegraphics[width=0.45 \textwidth]{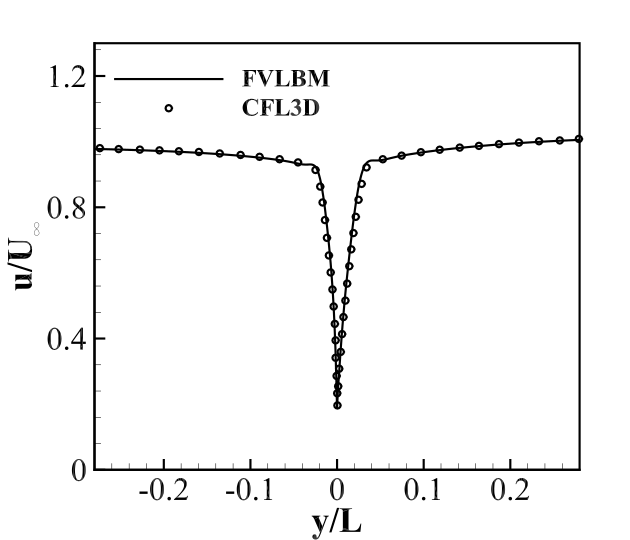}
           	}
 	\caption{\label{naca0012uxalpha3} The x-component of velocity at (a) $x/L = 0.0$, (b) $x/L = 0.25$, (c) $x/L = 0.5$, (d) $x/L = 0.75$, and (e) $x/L = 1$ for flow around NACA0012 airfoil with $\alpha = 3^{\circ}$.}
\end{figure}

\begin{figure}
 	\centering
 	\subfigure[]{
 			\includegraphics[width=0.45 \textwidth]{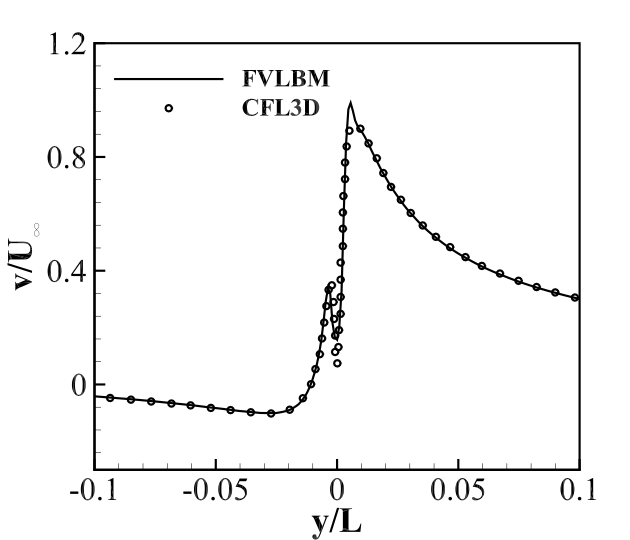}
 			}
    \subfigure[]{
     		\includegraphics[width=0.45 \textwidth]{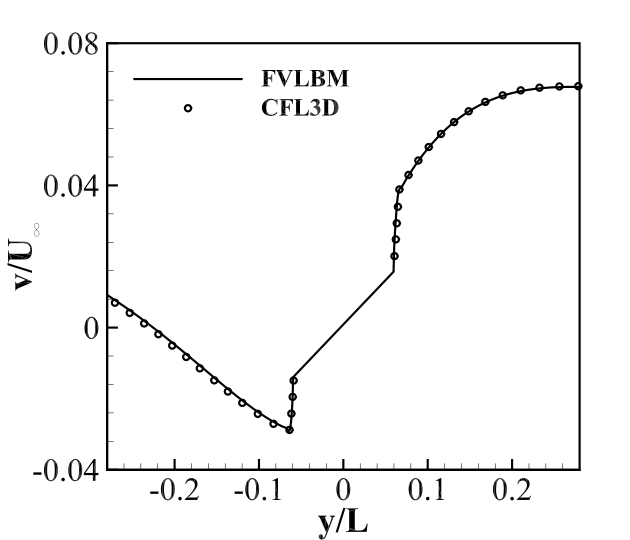}
         	}
    \subfigure[]{
        	\includegraphics[width=0.45 \textwidth]{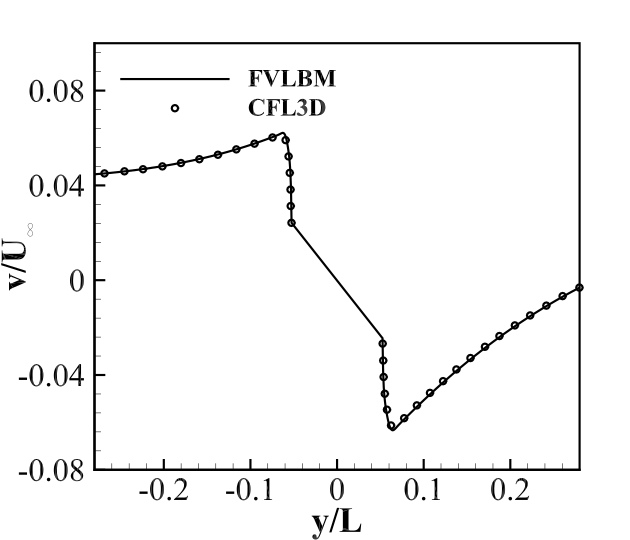}
          	}
    \subfigure[]{
            \includegraphics[width=0.45 \textwidth]{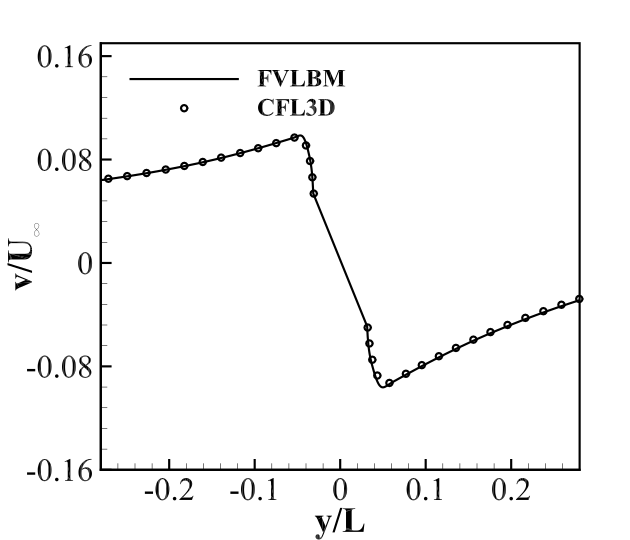}
           	}
    \subfigure[]{
            \includegraphics[width=0.45 \textwidth]{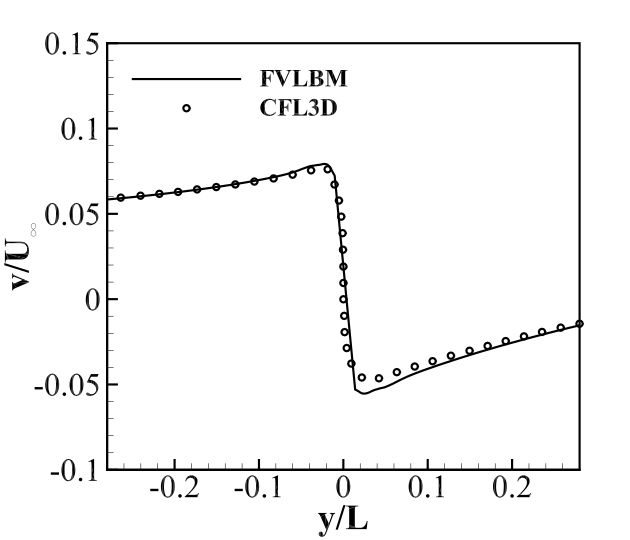}
           	}
 	\caption{\label{naca0012vxalpha3} The y-component of velocity at (a) $x/L = 0.0$, (b) $x/L = 0.25$, (c) $x/L = 0.5$, (d) $x/L = 0.75$, and (e) $x/L = 1$ for flow around NACA0012 airfoil with $\alpha = 3^{\circ}$.}
\end{figure}

\begin{figure}
 	\centering
 	\subfigure[]{
 			\includegraphics[width=0.45 \textwidth]{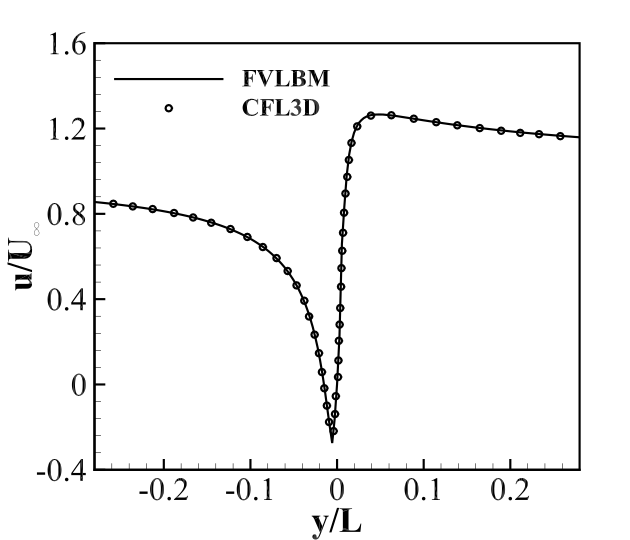}
 			}
    \subfigure[]{
     		\includegraphics[width=0.45 \textwidth]{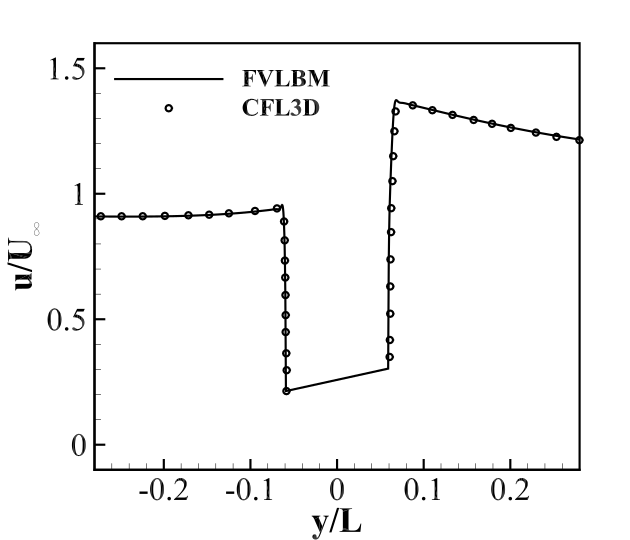}
         	}
    \subfigure[]{
        	\includegraphics[width=0.45 \textwidth]{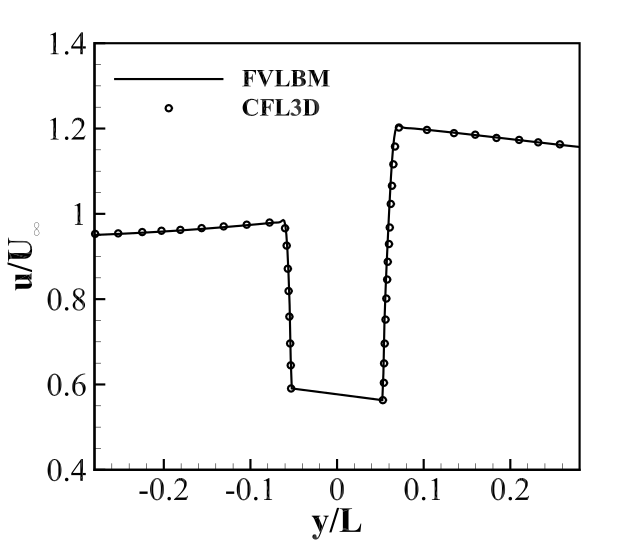}
          	}
    \subfigure[]{
            \includegraphics[width=0.45 \textwidth]{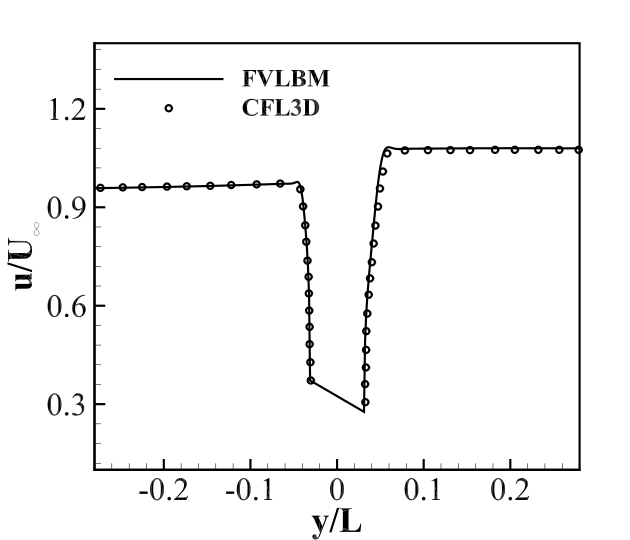}
           	}
    \subfigure[]{
            \includegraphics[width=0.45 \textwidth]{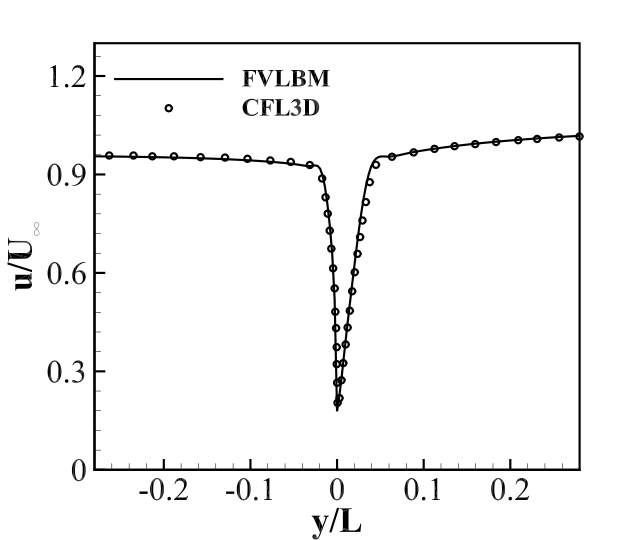}
           	}
 	\caption{\label{naca0012uxalpha7} The x-component of velocity at (a) $x/L = 0.0$, (b) $x/L = 0.25$, (c) $x/L = 0.5$, (d) $x/L = 0.75$, and (e) $x/L = 1$ for flow around NACA0012 airfoil at $\alpha = 7^{\circ}$.}
\end{figure}

\begin{figure}
 	\centering
 	\subfigure[]{
 			\includegraphics[width=0.45 \textwidth]{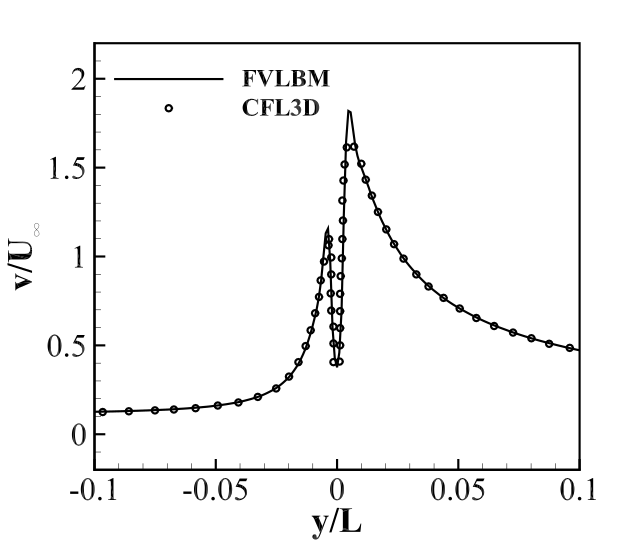}
 			}
    \subfigure[]{
     		\includegraphics[width=0.45 \textwidth]{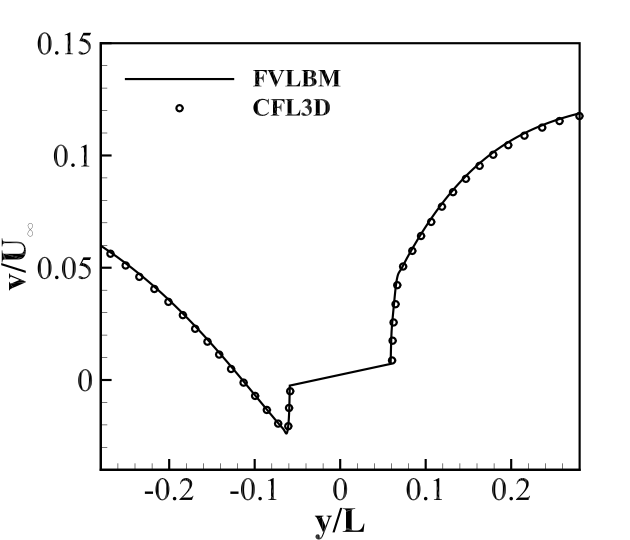}
         	}
    \subfigure[]{
        	\includegraphics[width=0.45 \textwidth]{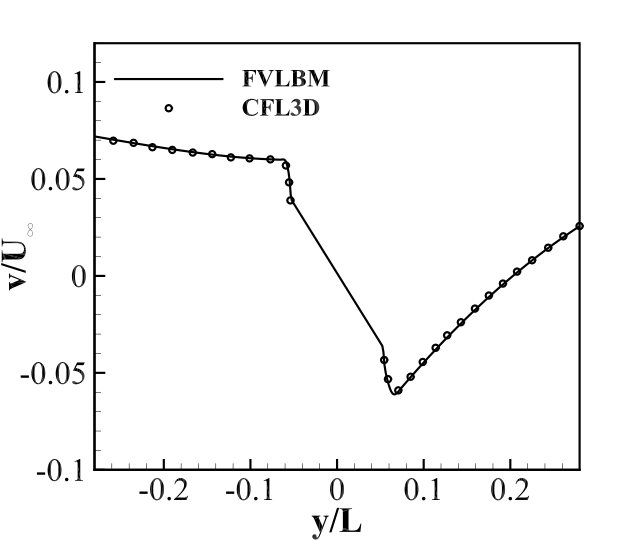}
          	}
    \subfigure[]{
            \includegraphics[width=0.45 \textwidth]{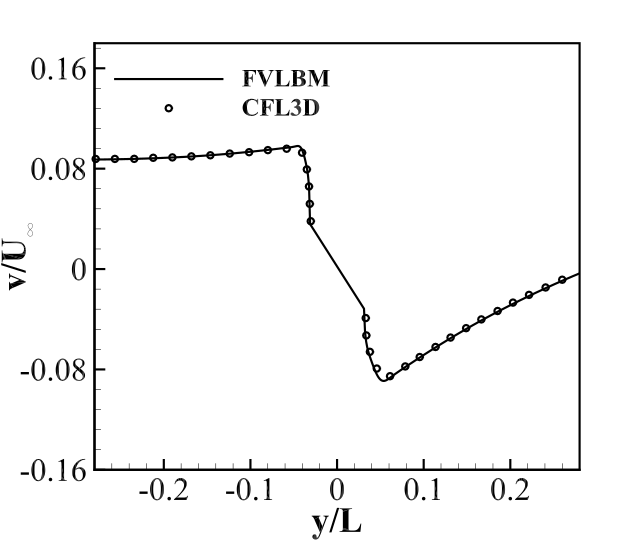}
           	}
    \subfigure[]{
            \includegraphics[width=0.45 \textwidth]{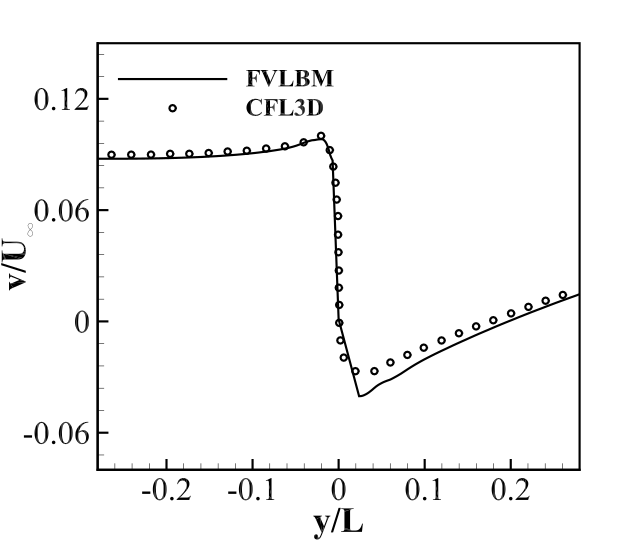}
           	}
 	\caption{\label{naca0012vxalpha7} The y-component of velocity at (a) $x/L = 0.0$, (b) $x/L = 0.25$, (c) $x/L = 0.5$, (d) $x/L = 0.75$, and (e) $x/L = 1$ for flow around NACA0012 airfoil at $\alpha = 7^{\circ}$.}
\end{figure}

\begin{figure}
 	\centering
 	\subfigure[]{
 			\includegraphics[width=0.45 \textwidth]{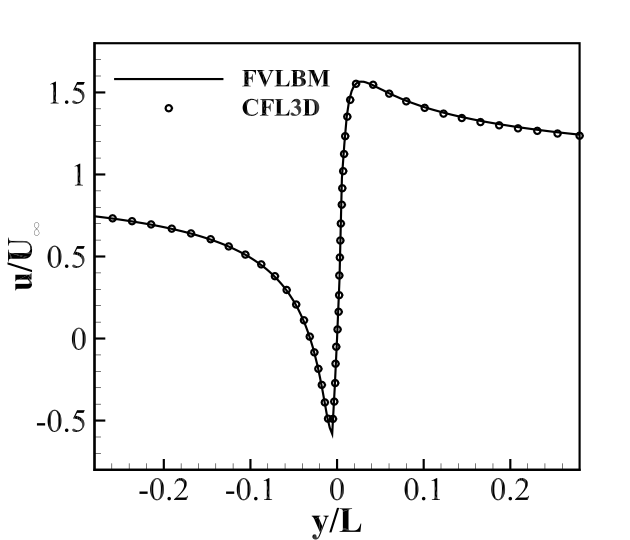}
 			}
    \subfigure[]{
     		\includegraphics[width=0.45 \textwidth]{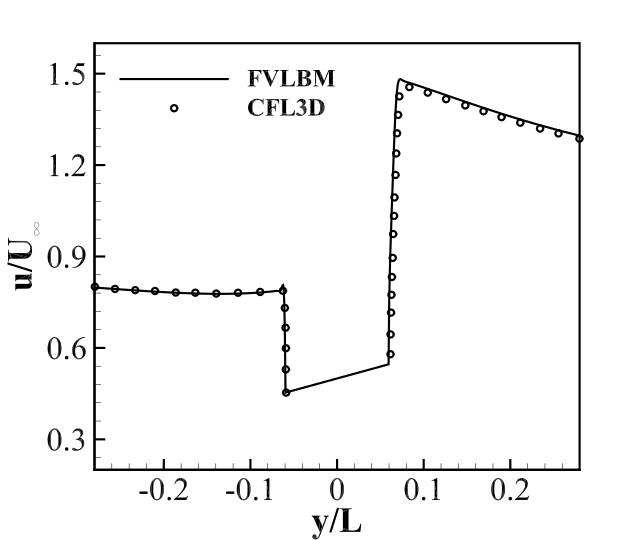}
         	}
    \subfigure[]{
        	\includegraphics[width=0.45 \textwidth]{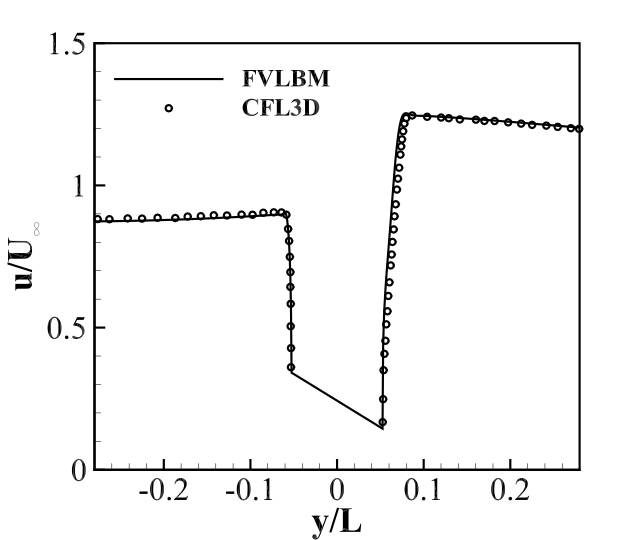}
          	}
    \subfigure[]{
            \includegraphics[width=0.45 \textwidth]{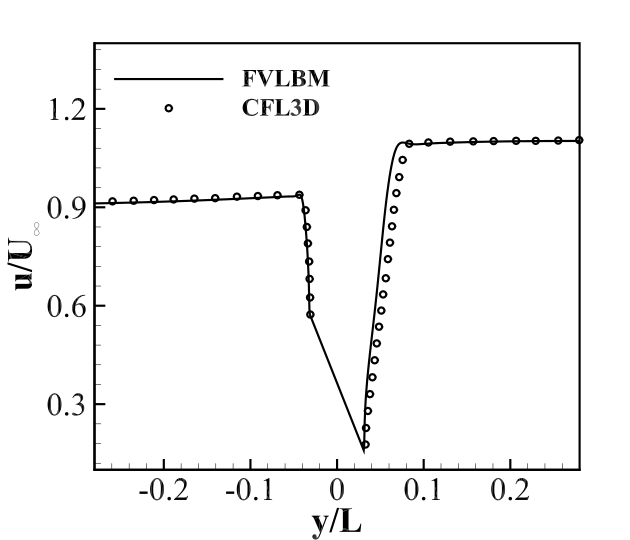}
           	}
    \subfigure[]{
            \includegraphics[width=0.45 \textwidth]{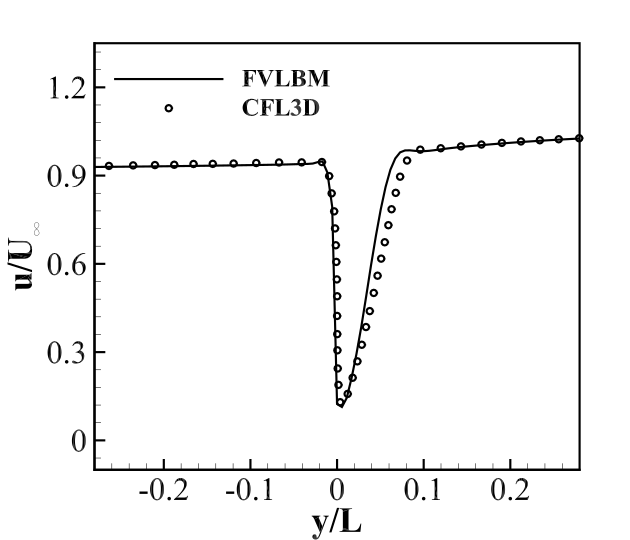}
           	}
 	\caption{\label{naca0012uxalpha12} The x-component of velocity at (a) $x/L = 0.0$, (b) $x/L = 0.25$, (c) $x/L = 0.5$, (d) $x/L = 0.75$, and (e) $x/L = 1$ for flow around NACA0012 airfoil at $\alpha = 12^{\circ}$.}
\end{figure}

\begin{figure}
 	\centering
 	\subfigure[]{
 			\includegraphics[width=0.45 \textwidth]{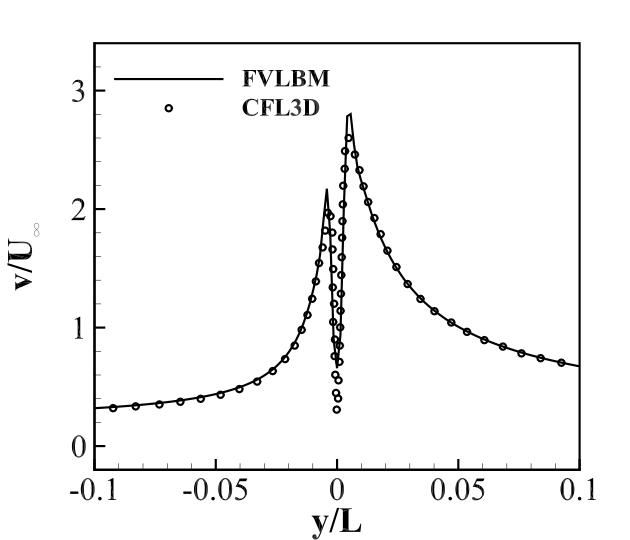}
 			}
    \subfigure[]{
     		\includegraphics[width=0.45 \textwidth]{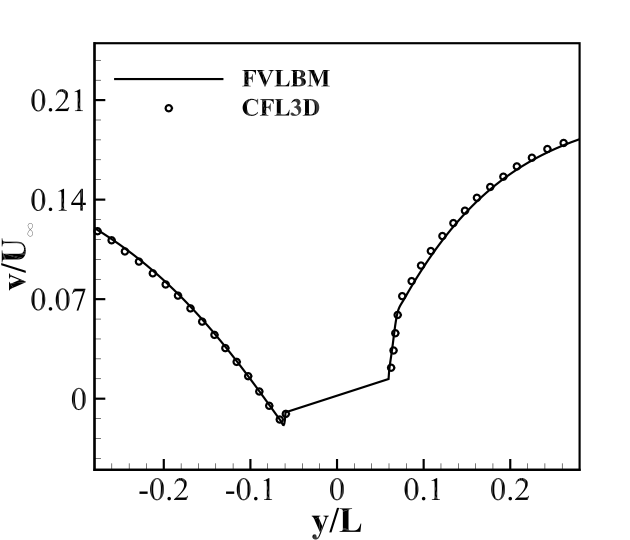}
         	}
    \subfigure[]{
        	\includegraphics[width=0.45 \textwidth]{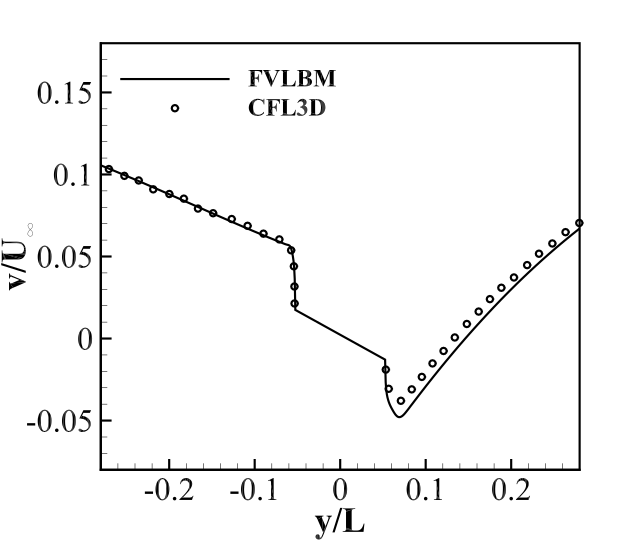}
          	}
    \subfigure[]{
            \includegraphics[width=0.45 \textwidth]{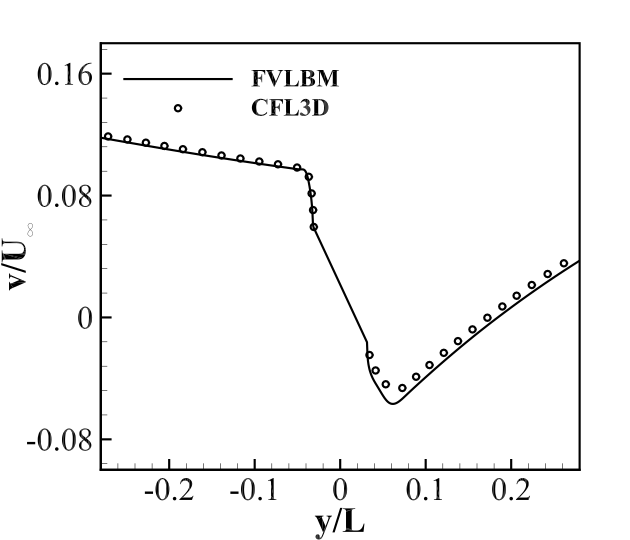}
           	}
    \subfigure[]{
            \includegraphics[width=0.45 \textwidth]{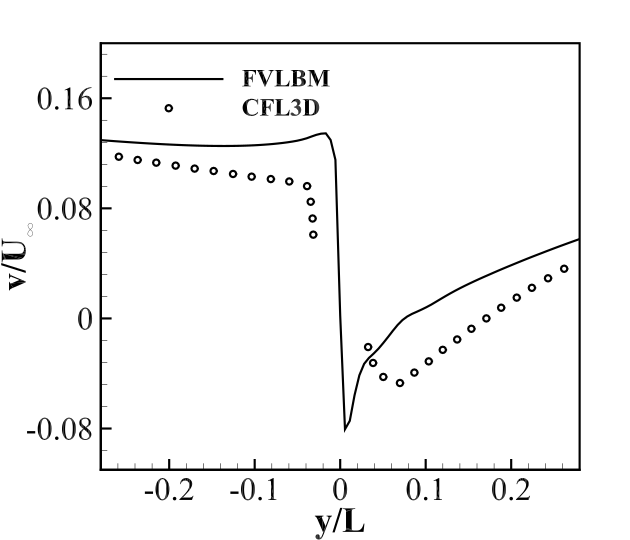}
           	}
 	\caption{\label{naca0012vxalpha12} The y-component of velocity at (a) $x/L = 0.0$, (b) $x/L = 0.25$, (c) $x/L = 0.5$, (d) $x/L = 0.75$, and (e) $x/L = 1$ for flow around NACA0012 airfoil at $\alpha = 12^{\circ}$.}
\end{figure}

Figs.~\ref{naca0012uxalpha0}-\ref{naca0012vxalpha12} show the x- and y-component velocity at five positions for four angles of attack. For $\alpha=0^{\circ}$, $3^{\circ}$ and $7^{\circ}$, good results are obtained compared with CFL3D data for both $u$ and $v$ velocity profiles. But at $\alpha=12^{\circ}$, as it close to stalling angle of attack, the differences between two results are increased, especially at trailing edge.

The lift coefficients $C_l$ and the drag coefficients $C_d$ for flow around NACA0012 airfoil are present in Table. \ref{tab:naca0012clcd}. For the lift coefficients, our results are higher than CFL3D data and for drag coefficients, our results are more closer to CFL3D data than Pellerin et al.'s.

\begin{table}
    \centering
	\caption{\label{tab:naca0012clcd} Lift and drag coefficients of NACA0012 airfoil at $Re = 10^5$}
	\begin{tabular}{ccccccc}
	  \toprule
	    \multicolumn{1}{c}{\multirow{2}{*}{AOA($^\circ$)}} & \multicolumn{3}{c}{$C_l$}& \multicolumn{3}{c}{$C_d$}   \\
	    \cline{2-4}
	    \cline{5-7}
	    ~& \multicolumn{1}{c}{CFL3D} & \multicolumn{1}{c}{Pellerin et al.\cite{pellerin2015implementation}} & \multicolumn{1}{c}{Present} &
	       \multicolumn{1}{c}{CFL3D} & \multicolumn{1}{c}{Pellerin et al.\cite{pellerin2015implementation}} & \multicolumn{1}{c}{Present}    \\
	  \midrule
	     0 & 0.0000 & 0.0000 & 0.0000 & 0.0128 & 0.0100 & 0.0112  \\
	     3 & 0.3237 & 0.3257 & 0.3297 & 0.0130 & 0.0104 & 0.0118  \\
	     7 & 0.7449 & 0.7457 & 0.7572 & 0.0157 & 0.0128 & 0.0151  \\
	    12 & 1.1809 & 1.1623 & 1.1222 & 0.0275 & 0.0263 & 0.0263  \\
	  \bottomrule
	\end{tabular}
\end{table}

\section{Conclusions}\label{Conclusions}
In this study, the original finite volume LBM presented in the first part of this paper is improved with the implement of implicit-explicit Runge-Kutta (IMEX) temporal discretization scheme. As treated with implicit method, the largest time step limited by the stability criterion of collision term is removed completely and the computational efficiency can be enhanced. A simple test case of lid-driven square cavity flow shows that the IMEX can decline the computational time about one order of magnitude compared with explicit Euler scheme. To model the effect of turbulence, the $k-\omega$ SST turbulence model is coupled into the present FVLBM scheme. For the test case of turbulent flow over the backward-facing step, the method used in this paper can capture the main features of large separation flow and numerical results are good agreement with experimental data. But the residual decline very slow, it means that the inlet and outlet boundary conditions used in this paper can not maintain the overall mass conservation, proper boundary conditions which easy to implement on unstructured grid need to further studies. For turbulent flow around the NACA0012 airfoil, good results also can be obtained compared with CFL3D data at relatively small amount of grid cells. Besides, the hybrid grid used in this test case shows again that the great flexibility of FVLBM for treatment the complex geometries. Finally, this coupling FVLBM scheme retain all the feature present in the first part of this paper.

\section*{Acknowledgements}
This work has been financially supported by National Natural Science Foundation of China (11472219), 111 project of China (B17037) and the Discovery Grant of the Natural Sciences and Engineering Research Council (NSERC) of Canada.

\clearpage
\bibliography{ref}

\end{document}